\documentclass{aa}  
\usepackage{graphicx}
\usepackage{lscape}
\usepackage{txfonts}
\DeclareUnicodeCharacter{223C}{~}
\usepackage{color}
\usepackage{soul}
\usepackage{lineno}
\defcitealias{2011Mosquera}{M\&K}

\begin{document} 
   \title{Microlensing time-scales and flux magnification probabilities of a sample of 204 lensed quasars} 

\titlerunning{Microlensing properties of a sample of 204 lensed quasars}
   \author{F. \'Avila-Vera \inst{1},
          V. Motta\inst{1}, E. Mediavilla\inst{2,3}
          }
   \institute{Instituto de Física y Astronomía, Facultad de Ciencias, Universidad de Valparaíso, Av. Gran Breta\~na 1111, Valparaíso, Chile.\\
\email{felipe.avilav@postgrado.uv.cl,veronica.motta@uv.cl },
     \and Instituto de Astrofísica de Canarias, Vía Láctea S/N, La Laguna 38200, Tenerife, Spain, \and Departamento de Astrofísica, Universidad de la Laguna, La Laguna 38200, Tenerife, Spain,
             }

   \date{Received --, --; accepted --,--}

  \abstract
{Quasar microlensing is both a very useful tool in cosmology and astrophysics, and a source of uncertainty in some studies like the determination of the Hubble constant from lensed quasars. Microlensing probability and time-scales have been statistically studied using as a reference scale the Einstein ring crossing time of an isolated mass.}
  {Our goal is to extend the statistical analysis of microlensing to all currently known lensed quasars with available data, considering realistic optical depths and the gravitational effect of the lens galaxy. We take into account new observational results about quasar sizes and peculiar velocities of lens galaxies.} 
  {We apply automatic lens modeling to the 204 systems available. For each image, we compute microlensing magnification maps and histograms.} 
  {Using thin disk source sizes scaled to take into account recent measurements of accretion disk sizes, we find a mean source crossing time of $2.59\pm 0.07$ years. The mean Einstein radius crossing time is $ 11.29 \pm 0.05$ years.  When a fraction of mass in microlenses $\alpha=0.2$ is adopted, we find a good matching between the modeled histogram of mean microlensing magnifications for the images in our sample and the experimental histogram of microlensing magnifications.}
  {From the modeling of microlensing magnification histograms, we estimate the average half-light radius of the quasar source, $R_{1/2}=5.4\pm 2.7$ light-days, and a lower limit to the mass fraction in microlenses, $\alpha\ge 0.15$. From the microlensing magnification maps, we find that a lensed quasar image has a mean probability of approximately 9\% of being involved in a high-magnification event ($\Delta m \le -0.32$). We select a group of images with the largest probabilities and the smallest crossing times.}
\maketitle \nolinenumbers

\section{Introduction}

\label{sec:introduction}
The discovery of the first gravitationally lensed quasar in 1979 by \citet{1979Walsh} opened a new field of study that had only been theorized. Nearly 45 years have passed since then, and the number of known systems has reached more than 300 \citep{1998Munoz,2018Treu,2018Lemon,2024Chan}. This number is continually growing with projections for about 10,000 more in the coming years \citep{2010Oguri,2015Collett,2019Ivezic}.

Multiply imaged quasars provide valuable insights into the mass distribution, formation, and evolution of lens galaxies, either disk or elliptical galaxies, the most common type of lens \citep{2010Treu}. They also help to constrain the stellar initial mass function and to measure various cosmological parameters \citep[see][for recent reviews]{2019JimenezVicente,2024Shajib}.  In particular, a notable capability of gravitationally lensed quasars is their potential to determine the Hubble constant (H$_0$). Incorporating this method into the current suite of measurements may help to resolve the persistent $5\sigma$  tension in the inferred value of the Hubble constant \citep{2023Tully,2021DiValentino,2024Verde}.

In general, the macrolens responsible for producing multiple images of a quasar is well described by the smooth gravitational potential of the lens galaxy. However, real galaxies are granular: stars (and possibly other compact objects) introduce small-scale perturbations to the potential that differentially magnify distinct quasar emitting regions. This effect is called gravitational microlensing and produces time- and wavelength-dependent variations in the image flux ratios \citep[e.g.][]{1997Refsdal,2006Wambsganss,2024Vernardos}. Quasar microlensing is therefore a powerful probe of quasar structure across a wide range of spatial scales, from the immediate vicinity of the Supermassive black hole (SMBH) event horizon \citep{2010Morgan}, to the accretion disk \citep{2020Cornachione,2021Fian}, the broad-line region \citep[BLR;][]{2002Abajas,2024Hutsemekers}, and the dusty torus \citep{2013Sluse}. At the same time, microlensing statistics constrain the fraction of mass in compact objects and can be used to test lens-galaxy stellar mass functions (e.g., Present-Day Mass Function/Initial mass Function) and the abundance of potential dark-matter candidates \citep[e.g.][]{2001Wyithe,2019JimenezVicente,2023EstebanGutierrez,2023Awad}.

The advent of wide-field time-domain surveys, most notably the Vera C.\ Rubin Observatory Legacy Survey of Space and Time (LSST; \citealt{2019Ivezic}), will deliver multi-band monitoring for large samples of lensed quasars and will make it feasible to systematically study high-magnification events (HMEs). These phenomena are typically defined as sharp, transient brightenings caused by the source approaching or crossing a micro-caustic, and they can provide especially strong constraints on the source size and surface-brightness profile because the caustic effectively ``scans'' the emitting region \citep{2008Anguita,2015Mediavilla}. In this context, dedicated forward-modeling tools tailored to LSST cadences have been developed \citep{2020Neira}, and recent simulations predict that of order $\sim 60$ HMEs per year (with amplitudes $>0.3$ mag in $r$) could be detectable based on simulated $\sim~$4 billion light curves \citep{2025Neira}. These forecasts motivate early-warning and triggering strategies for dense follow-up during HMEs, when the information content for quasar-structure inference is maximal \citep[e.g.][]{2024Fagin}.

Consequently, the study of microlensing properties of known lensed quasars, in particular of its time-scale flux variations, is a key ingredient for the analysis of future systems. Early studies \citep[see, e.g., ][]{2004Kochanek,2006Wambsganss}
indicate that the microlensing time-scale is between months and years. In an extensive analysis  \cite{2011Mosquera} (hereafter, \citetalias{2011Mosquera}) developed a procedure to estimate this time scale for the 87 known systems at that time, with timescales ranging from months to decades. \citetalias{2011Mosquera} based their analysis on the Einstein radius ($\theta_E$) crossing time, a quantity defined for a single, isolated lens, which does not take into account the cooperative effects between microlenses corresponding to the relatively large optical depths typical of lensed quasars and the external gravitational field of the lens galaxy.

For compact microlenses, microlensing magnification patterns are scaled to the Einstein radius of the microlenses in the sense that if we, for instance, change the mass of the microlenses, $M\to M'$, all the lengths should be re-scaled by a $\sqrt{M'/M}$ factor \citep[mass-length invariance,][]{2006Wambsganss}. However, apart from this global change of scale, the Einstein radius crossing time may not be informative about the time scales of flux variations or about any pseudo-periodical properties of quasar microlensing \citep{2002Wyithe}. The frequency of variations is dominated by the optical depth of the microlenses and the shear of the macroscopic gravitational field of the lens galaxy \citep{1986Paczynskib,2005GilMerino}. On the one hand, even at low optical depth,  closed caustic curves (the typical astroids) are magnified by the effect of the shear. On the other hand, as the optical depth increases, single closed caustic curves can progressively interconnect forming first constellations and then intricate networks with high spatial frequencies (compensated by large demagnification regions).  None of these effects are directly related to the Einstein radius crossing time for an isolated particle \cite{1992Schneider}. The way to obtain more information about the properties of microlensing variability is, then, to simulate microlensing for the particular conditions of the gravitational field at each lensed quasar image.  Microlensing magnification maps are the usual tool applied in this kind of studies \citep{1986Kayser}.

As mentioned before, nearly fifteen years have passed since the work by \citetalias{2011Mosquera}, and the number of known lensed quasar systems has almost quadrupled \citep{2024Chan}. 
Moreover, new information about the size of the quasar accretion disk and about the lens galaxy peculiar velocities, both critical to study the microlensing time scales, are now available. Specifically, microlensing \citep{2012JimenezVictente,2020Cornachione} and reverberation mapping measurements \citep{2018Fausnaugh,2022Jha} indicate that the accretion disk size is several times larger than the estimates based on the thin disk model by \citep[][hereafter, S\&S]{1973Shakura}. Taking this effect into account, we rescale the theoretical size previously used in \citetalias{2011Mosquera} using microlensing estimations.

On the other hand, in addition to the calculations based on Einstein ring crossing-times, and to take into account realistic values of the optical depth of lensed quasar images, we calculate magnification maps for every image, which also informs us about the relative probability (and time-scales) of different levels of microlensing activity. To compute the statistically significant magnification probabilities, we are going to use fast multipole method–inverse polygon mapping \citep[FMM–IPM;][]{2022JimenezVicente}, which makes feasible the calculation of large enough microlensing magnification maps for a large number of lensed images. As mentioned above, the amplitude of microlensing flux magnification is another important property that we will analyze from these magnification maps.

 As a first step in this study, we perform homogeneous and simple parametric automatic modeling for all the gravitational lens quasars to avoid systematics associated with the lens modeling process. Consequently, we also obtain a complete sample of lens model parameters useful to analyze the properties of the population of lensed quasars systems.

The paper is organized as follows. Section 2 presents our data set. Section 3 describes the macro-modeling procedure, the computation of magnification maps, and the calculation of relevant timescales and probabilities. Results are presented in Section 4 and discussed in Section 5. Finally, the main conclusions are summarized in Section 6. Throughout this paper, we adopt the following cosmological
parameters: $\Omega_{m} = 0.3$,  $\Omega_{m} = 0.7$,
$\rm H_{0}~=~70~km/s/Mpc$. 

\begin{figure}
        \centering
        \includegraphics[width=1\linewidth]{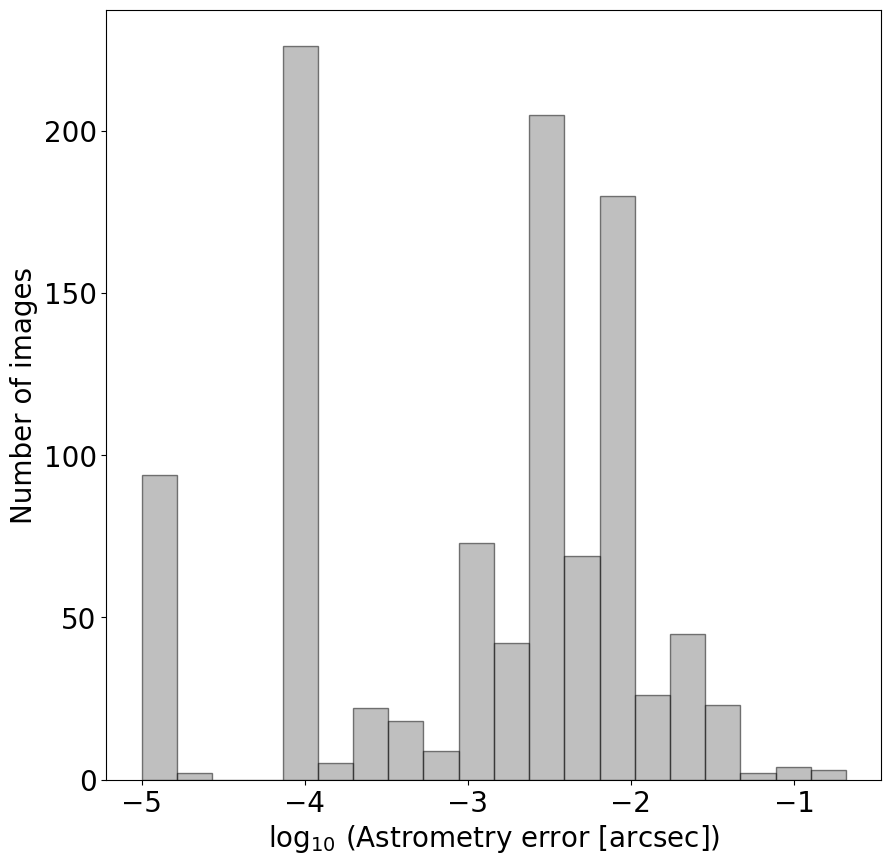}
        \caption{Astrometric error distribution for the sample of images modeled.}
        \label{fig:error_astrometri}
\end{figure}

\section{Data}

The first uniform and high-resolution image database of gravitationally lensed quasars was provided in the 1990s by the CfA–Arizona Space Telescope LEns Survey\footnote{\url{https://lweb.cfa.harvard.edu/castles/}}] \citep[CASTLES;][]{1998Munoz}. Using the Hubble Space Telescope (HST), broad-band images were obtained for $\sim 100$ systems, primarily in the filters F160W, F814W, and F555W, including 15 galaxy-scale quadruply lensed quasars.

Decades later, in 2015, the Strong Lensing Insights into the Dark Energy Survey \citep[STRIDES;][]{2018Treu} initiated a new search for lensed quasars in the Dark Energy Survey \cite{2005TheDarkEnergySurveyCollaboration}, aiming to discover new quadruply lensed quasars and to select the best candidates for time-delay follow-up and H$_{0}$ estimation. The initial search was subsequently expanded to incorporate Gaia data \citep{2018KroneMartins}, which enabled the exclusion of stars as contaminants \citep{2017Lemon,2018Agnelloa,2023Lemon}.

A few years ago, the Gravitationally Lensed Quasar Database (GLQD) listed the names and references of most known lensed quasars. However, the information provided for the systems was not homogeneous, and the page has since been taken down. New projects, such as the Strong Lensing Database (SLED\footnote{\url{https://sled.amnh.org/}}), have emerged to fill this gap, aiming to provide a unified and comprehensive database of strong lensing systems. However, it still does not provide complete information (e.g., photometry in multiple bands for the images) for all the known systems.

Considering this inhomogeneity, we construct a dataset to identify the largest collection of systems containing photometric information for the images, relative astrometry, and the redshifts of both the lens and source. To achieve this, we begin our search with the systems published in the CASTLES database, where objects are documented with photometric data in the HST bands, relative astrometry, and redshift information for the source and lens.
Next, we turn to the GLQD and search the literature for photometric and relative astrometric data on the latest discovered systems. We also include systems not listed in either of the databases above. Since the redshift estimates for the source and lens are not always reported in the same publication, we search for the most recent publicly available values.
Our search produces as output a sample of 314 known systems, each with different levels of information; this census can be used following the example \footnote{\url{https://github.com/felavila/qumas/blob/main/examples/how_to_use_census.ipynb}}

Later in the paper, we will compare our estimates of microlensing
magnification probabilities with the observed microlensing histogram
presented by \citeauthor{2009Mediavilla} (\citeyear{2009Mediavilla}, hereafter MED09; see also \citealt{2024Mediavilla}). This empirical histogram is
constructed from single‑epoch microlensing measurements of image pairs
and, therefore, does not preserve the sign of the magnification. For
this reason, we also compute probability distributions for the unsigned
value of the microlensing magnification.

\begin{figure*}
    \centering
    \includegraphics[width=1.0\textwidth]{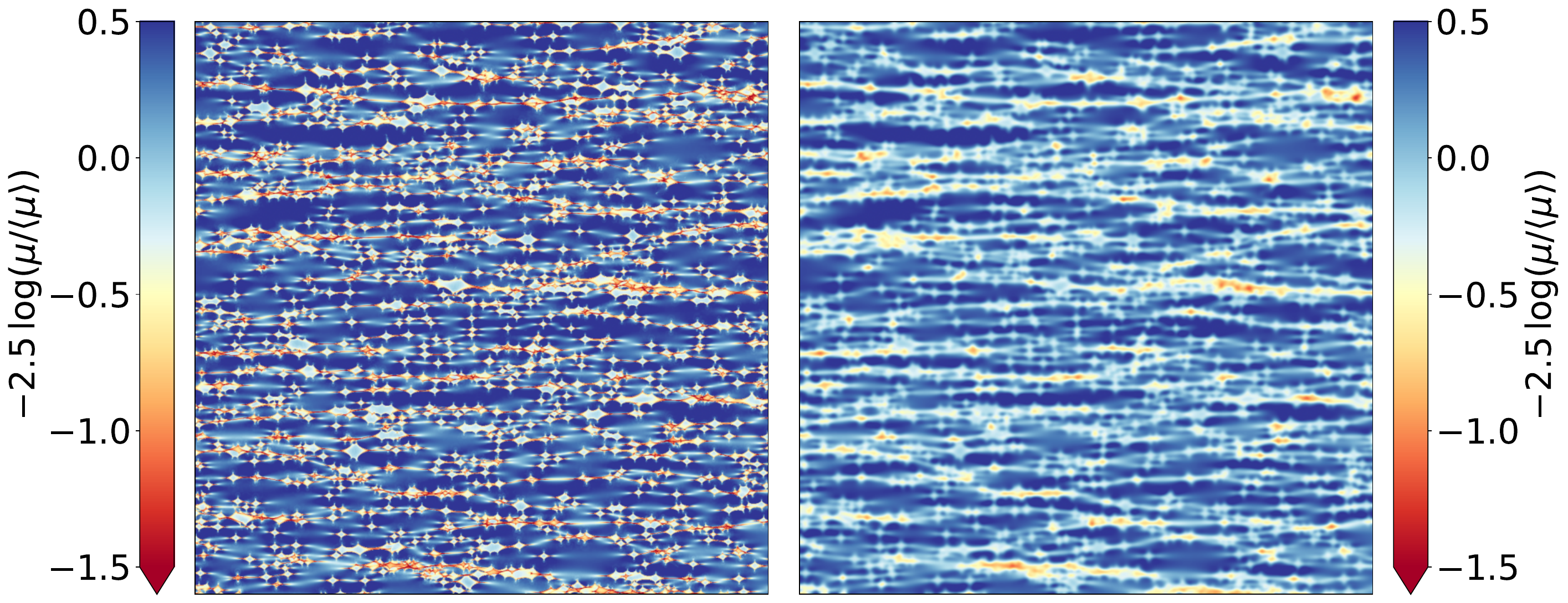}
    \centering
    \caption{Microlensing magnification patterns produced by stars in the lensing galaxy corresponding to image~A of the system DES2158$-$5812. \textbf{Left:} Original magnification map. The color scale encodes different magnification levels, while the diamond-shaped curves trace the caustics.\textbf{Right:} The same map after convolution with a Gaussian profile with a sigma size of 3.72 [pix], showing the effective magnification pattern. Both panels have a size of $1200 \times 1200$ pixels with a pixel size of 1.1 [pix/light-days] and 60 R$_{\mathrm{E}}$. The maps were produced using the code developed in \citet{2022JimenezVicente}.
    }
    \label{fig:mapa_magnification}
\end{figure*}

\section{Methods}
\subsection{Macro-lens modeling}
Over the last decades, different lens modeling software \citep{2013Lefor} with a variety of approaches has been applied to study individual systems \citep{2022Shajib}, meanwhile, the use of machine learning techniques is starting to appear to tackle the upcoming thousands of objects that will be discovered \citep{2025Andika,2025Erickson}. For example, to model complex systems using high-resolution images (i.e., quadruply lensed quasars with extended arcs), several codes exist, in particular the two well-tested modeling codes \texttt{Lenstronomy} \citep{2015Birrer,2018Birrer} and \texttt{Glee} \citep[Gravitational Lens Efficient Explorer;][]{2010Suyu}. Both codes were used in auto-modeling task \citep[see][]{2023Schmidt,2023Ertl}, concluding that for some of the systems,  human supervision is still required. In fact, few of the modeling codes offer a native automatic process. An example is \texttt{AutoLens} \citep{2021Nightingale}. This code uses a non-linear search to determine the set of light and mass profile parameters that best fit the data using images of the objects (see more details in \texttt{pyautolens} docs \footnote{\url{https://pyautolens.readthedocs.io/}}). 

Another difficulty is that the kind of observational data required by the most complex/sophisticated codes (for example, line-of-sight velocity dispersion, high-resolution images \citealt{2025Knabel}) is not available for a significant part of the known quasar lenses. 

As we aim to study as many systems as possible using a homogeneous procedure, we adopt a straightforward approach based solely on photometric data from the lensed quasar images and on the relative positions between the images and the lens galaxy, which are usually available in the literature. We selected \texttt{Lensmodel} \citep{2001Keetonb,2011Keeton}, a parametric gravitational lens modeling code that the community has extensively tested for over two decades. This code supports a wide range of analytic mass distributions and can fit lens systems using constraints from image positions and flux ratios. Modeling can be performed either in the image plane or in the source plane. In our case, we adopt the image-plane approach, in which the code minimizes a standard chi-square statistic of the form:

\begin{equation}
\label{eq:chi2_lensmodel}
\chi^2 = \sum_i \frac{|\theta_{i,\mathrm{mod}} - \theta_{i,\mathrm{obs}}|^2}{\sigma_{i,\theta}^2} + \sum_i \frac{(f_{i,\mathrm{mod}} - f_{i,\mathrm{obs}})^2}{\sigma_{i,f}^2},
\end{equation}

\noindent
where $\theta_{i,\mathrm{mod}}$ and $\theta_{i,\mathrm{obs}}$ are the modeled and observed image positions (in arcseconds) respectively, with positional uncertainties $\sigma_{i,\theta}$, and $f_{i,\mathrm{mod}}$ and $f_{i,\mathrm{obs}}$ are the modeled and observed fluxes with flux uncertainties $\sigma_{i,f}$. The goal is to determine the set of lens model parameters that minimizes $\chi^2$, thereby yielding the best match between the model predictions and the observational constraints.

Building on this well-established code, we developed \texttt{QuMAS} (Quasar Microlensing Analysis\footnote{\url{https://github.com/felavila/qumas}}), a Python wrapper that implements an automatic modeling pipeline and the methodologies presented in this work. This framework enables fully automated lens modeling, allowing us to systematically analyze a large number of systems with a consistent methodology. Below, we describe in detail the steps of this automatic pipeline. In Section~\ref{ss: Comparison with previous automatic modeling studies}, we compare our results with those from more complex or sophisticated models to assess how much a single model that captures the essence of the phenomenon, can differ from a more elaborated one.

Next, we will explain our automatic modeling procedure and its application to our current data sample. The steps are as follows:

We start by modeling the 314 systems using the most recent and accurate photometry and astrometry available in the I band ($\lambda = \rm 8140$~\AA) or similar. 
The first filter checks whether the system contains known relative astrometry and photometry for all images, which results in a sample of 204 systems that can be modeled.
The initial model applied is a singular isothermal sphere \cite[SIS,][]{2001Keetona}, the simplest mass distribution and a common first approximation in many cases. The initial conditions for image uncertainties are derived from the data census. The galaxy position uncertainty is fixed at $10^{-3}$~arcseconds, providing a strong constraint, and the fluxes are also taken from the census with an associated error of 20\%. This uncertainty in the fluxes accounts for the effects of microlensing in one or more images, which can introduce additional magnification. The value is chosen as a conservative estimate based on the literature \citep{2010Morgan,2016VivesArias,2018Morgan}, where reported uncertainties range between 10\% and 50\%. 

To evaluate the convergence of the model, besides a $\chi^{2}$ value, we additionally assess the agreement between the model and the data by computing the maximum separation between the calculated and the observed image positions($\Delta\theta_{\max}$), defined as follows:
    \begin{align}
    \label{eq:max_theta}
    \Delta\boldsymbol{\theta}_{i} &= \boldsymbol{\theta}_{i,\mathrm{mod}} - \boldsymbol{\theta}_{i,\mathrm{obs}},\\
    \Delta\theta_i &= \left|\Delta\boldsymbol{\theta}_{i}\right|,\\
    \Delta\theta_{\max} &= \max \Delta\theta_i,
    \end{align}
 This choice was motivated by the potential uncertainties associated with flux measurements and our threshold for determining the quality of the modeling. We consider a model successful if $\Delta\theta_{\max}$ $\leq$ $10^{-2}$~arcseconds. This threshold is based on the error distribution of the census images (Figure~\ref{fig:error_astrometri}). While the median positional error is $10^{-3}$~arcseconds, we adopt a threshold five times larger to allow for a simpler model.

If the model is successful, we take the values of the convergence ($\kappa$), which represents the dimensionless surface mass density of the lens; the shear ($\gamma$), which is associated with the distortion induced by the tidal gravitational field of the lens and its environment; and the Einstein radius, given its proportionality to the lens galaxy mass, as provided by \texttt{Lensmodel}. We then estimate the uncertainties using the Markov Chain Monte Carlo (MCMC) method implemented in \texttt{Lensmodel}. The MCMC is initialized using the modeling parameters as Gaussian priors, and is run with ten chains and a maximum of 10{,}000 steps.

If the condition ($\Delta \theta_{\max} \leq 10^{-2}$~arcseconds) is not satisfied, the system will be re-modeled based on the number of images. For doubles, we consider models such as singular isothermal ellipsoid \cite[SIE,][]{1994Kormann}, or SIS with $\gamma$, where shear represents an external gravitational perturbation introducing ellipticity. For quads, the modeling options expand to include SIE, SIS+$\gamma$, power-law mass distributions \cite[POW,][]{2001Keetona}, SIE+$\gamma$, or POW+$\gamma$. The larger number of images in quads allows for testing a greater variety of mass distributions due to the increased number of free parameters. 

If the system cannot be successfully modeled, we return to step 3 and modify the initial conditions. We first increase the images flux error to 50\%, keeping the lens galaxy position error unchanged. Next, we set the flux error to 20\% and increase the lens galaxy position error to 0.1~arcseconds. These changes help address scenarios where external shear effects are insufficient to characterize distortions, and the centroid of the potential does not coincide with the center of the light distribution of the lens galaxy. Finally, we attempt a model with a flux error of 50\% and a lens galaxy position error of 0.1 arcseconds.

If no modeling option succeeds after all these attempts, and to preserve the maximum number of systems in the sample, we select the model with the smallest $\Delta \theta_{\max}$ and \texttt{Lensmodel} $\chi^2$.

\subsection{Microlensing time-scales \label{sec:micro_time}}
    The relative motion between the observer, lens, and source results in the time-dependent microlens effect. Two temporal scales can be associated to microlensing \citep[for a recent review see Section 3.5.6 of][ ]{2024Vernardos}. One is the  basic temporal scale standard,  t$_{E}$, which is defined as the time it takes for the source to cross the Einstein radius of a single, isolated microlens (R$_{E}$) and can be written as:
     \begin{equation}
        \label{eq:te}
         t_E=\frac{R_E}{v},
    \end{equation}
    where $v$ corresponds to  the effective velocity of the source, and $R_{E}$ is, defined as:
    \begin{equation}
        R_E=D_{OS}\left[ \frac{4G M_{\star}}{c^2}\frac{D_{LS}}{D_{OL}D_{OS}}\right]^{1/2},
        \label{eq:Re}
    \end{equation}
   where $D_{OS}$, $D_{LS}$ and $D_{OL}$ correspond to the angular diameter distances between obs-source, lens-source, obs-lens, G is the gravitational constant, c is the speed of light and  $M_{\star}$ correspond to the mass of the star that is acting as a microlens. To calculate $R_E$, we assume a value of $M_{\star} = 0.3 M_{\odot}$. Two reasons drive this assumption: 1) this is the value estimated by \cite{1998Holtzman}  observing the Baade's window in the Galactic bulge using Hubble space telescope, 2) this value has been used largely in previous microlensing studies  \citep [see][] {2012Mao,2014JimenezVicente, 2018Fian,2022Paic}.

According to \citetalias{2011Mosquera}, the $t_{E}$ time scale is expected to be of the order of decades \citep[see][]{2009Smith,2017Bonaca}. However, if the source crosses  a compact region of very high magnification like one of the  caustic curves \citep{1992Blandford} shown in Figure \ref{fig:mapa_magnification}, it is possible to observe microlensing variations in shorter time scales; this time scale is known as the crossing time scale $t_{S}$ and is defined as
    \begin{equation}
        \label{eq:ts}
        t_S=\frac{R_S}{v}.
    \end{equation}
Here, $R_{S}$ represents the radius of the light source, and it can be estimated assuming a simple thin-disk model \citep{1973Shakura}, with a temperature profile of $T \propto R^{-3/4}$ thus \citepalias{2011Mosquera}, 

\begin{equation}
\label{eq:Rs}
\text{\fontsize{9.1pt}{10pt}\selectfont$
R_{S} = \frac{3.4\times10^{15}}{\sqrt{\cos i}}\,
\frac{D_{OS}}{r_H}\,
\left(\frac{\lambda}{\mu\mathrm{m}}\right)^{3/2}
\left(\frac{zpt}{3631\,\mathrm{Jy}}\right)^{1/2}
10^{-0.2\,(m-19)}\, h^{-1}\,\mathrm{cm}
$},
\end{equation}

\noindent 
where $i$ is the average inclination of the quasar disk, $zpt$ is the zero point, which depends on the photometric system and broad-band used, $\lambda$ is the reference wavelength corresponding to the mean of the band in which the observations were made, and $m$ is the source aparent magnitude after removing the gravitational lens magnification (i.e., demagnified magnitude). The values for $zpt$ and $\lambda$ are not always available in the literature, so we obtained them from the Filter Profile Service \citep{2020Rodrigo}\footnote{\url{https://svo2.cab.inta-csic.es/svo/theory/fps3/}} using the specified instrument and band. For objects where this information is not explicitly provided, we assume a Vega system $zpt$.

Following other authors, we adopt $i = \pi/3$, which corresponds to the average inclination of known systems \citep[see][]{2017Wildy}. Since not all systems have observations at the same wavelength ($\lambda_{obs}$), we rescaled all the obtained $R_{S}$  using the thin-disk model approximation, $R_{\lambda} \propto \lambda^{4/3}$, to a reference rest-frame wavelength of $\lambda = 0.250~\mu$m. Following \cite{2010Morgan}, we use the equation:

\begin{equation}
    R_{\lambda} = R_{S} \cdot \left( \frac{(1+z_s)\cdot\lambda}{\lambda_{obs}}\right)
   ^ {\left(4/3\right)},
    \label{eq:rescale}
\end{equation}

\noindent
where $R_{\lambda}$ corresponds to the size of the radius at restframe $\lambda$, and $R_{S}$ is the radius obtained from equation \ref{eq:Rs} and $\lambda_{obs}$ the wavelength of the observed band.

Now we introduce the velocity model that we will use to estimate the scales (see Equation~\ref{eq:te} and \ref{eq:ts}), also known as the effective velocity of the source ($v$) which is calculated considering the relative motions of observer, lens, and source \citep{1986Kayser} and is calculated using Equation~8 from \cite{2016Mediavilla}:

\begin{equation}
    v = \sqrt{V_1 + V_2 + V_3 + V_4} \label{eq:v_main},
\end{equation}
From where: 
\begin{subequations}
\label{eq:v}
\begin{align}
V_1 &= \left( \frac{v_{\mathrm{CMB}}}{1+z_{L}} \frac{D_{LS}}{D_{OL}} \right)^2, \label{eq:v_T1} \\
V_2 &= \left( \frac{\sqrt{2}\,\sigma_{\mathrm{pec}}(z_{L})}{1+z_{L}} \frac{D_{OS}}{D_{OL}} \right)^2 \label{eq:v_T2}, \\
V_3 &= \left(  \frac{\sqrt{2}\,\sigma_{\mathrm{pec}}(z_{S})}{1+z_{S}} \right)^2, \label{eq:v_T3} \\
V_4 &= 2\left( \frac{\sqrt{2}\,\sigma_{*}}{1+z_{S}} \frac{D_{OS}}{D_{OL}} \right) \label{eq:v_T4}.
\end{align}

\end{subequations}

In Equation~\ref{eq:v_T1}, $v_{\mathrm{CMB}}$ represents the projection of the CMB dipole velocity. This velocity depends on both the velocity and coordinates of the CMB, as well as the coordinates of the object. It is calculated following \cite{2010Poindexter}, assuming a velocity of CMB of $369.82 \pm 0.11 \ \rm{km/s}$ at coordinates RA $167^{\circ}.942 \pm 0^{\circ}.007$ and DEC $-6^{\circ}.944 \pm 0^{\circ}.007$ \citep{2020PlanckCollaboration}.

The peculiar velocity ($\sigma_{\mathrm{pec}}$) in Equations~\ref{eq:v_T2} and \ref{eq:v_T3} refers to the velocity at which a galaxy deviates from the Hubble flow at a given redshift. We estimate it,  following the steps of \cite{2016Mediavilla}, assuming the linear regime in a $\Lambda\mathrm{CDM}$ cosmology. In this framework, $\sigma_{\mathrm{pec}}$ can be expressed in terms of the cosmological growth rate factor $f(z)$ as:

\begin{equation}
\sigma_{\mathrm{pec}}(z) = \frac{\sigma_{\mathrm{pec}}(0)}{(1+z)^{1/2}} \frac{f(z)}{f(0)}, \label{velocidadpeculiar}
\end{equation}
\noindent
we adopt the peculiar velocity at redshift zero from \cite{2016Hess}~$\sigma_{\mathrm{pec}}(0) = 439 \pm 69\ \mathrm{kms^{-1}}$, and $f(z)$  can be calculated using:  

\begin{equation}
\label{eq:growt_rate}
f(z) = \Omega_{m}^{\gamma}(z),
\end{equation}

\noindent
here $\Omega_{m}(z)$ is the matter density parameter and $\gamma$ is the growth index, with $\gamma = 0.55$ for $\Lambda\mathrm{CDM}$ \citep{2022Avila}.

\noindent
Finally, $\sigma_{*}$ in Equation~\ref{eq:v_T3} represents the velocity dispersion of the stars in the lens galaxy. It is estimated using the SIE mass distribution, following previous works \citep{2008Grillo,2009Treu}:

\begin{equation} 
\sigma_{*}=\sqrt{\frac{ c^2 D_{OS} \theta_{E}}{4\pi D_{LS}}}, 
\label{eq} 
\end{equation}

\noindent
In those cases where the lens or source redshift is unknown, we follow \cite{2009Coe,2012Redlich,2020Robertson} and similar works, where they are typically assumed to be $z_{L} \sim 0.5$ and $z_{S} \sim 2.0$, respectively. This assumption allows us to be consistent with previous works and the distribution of redshift in our sample \cite[e.g. ;][]{2018Treu,2023Schmidt}

\begin{figure}[h]
    \centering
    \includegraphics[width=1\linewidth]{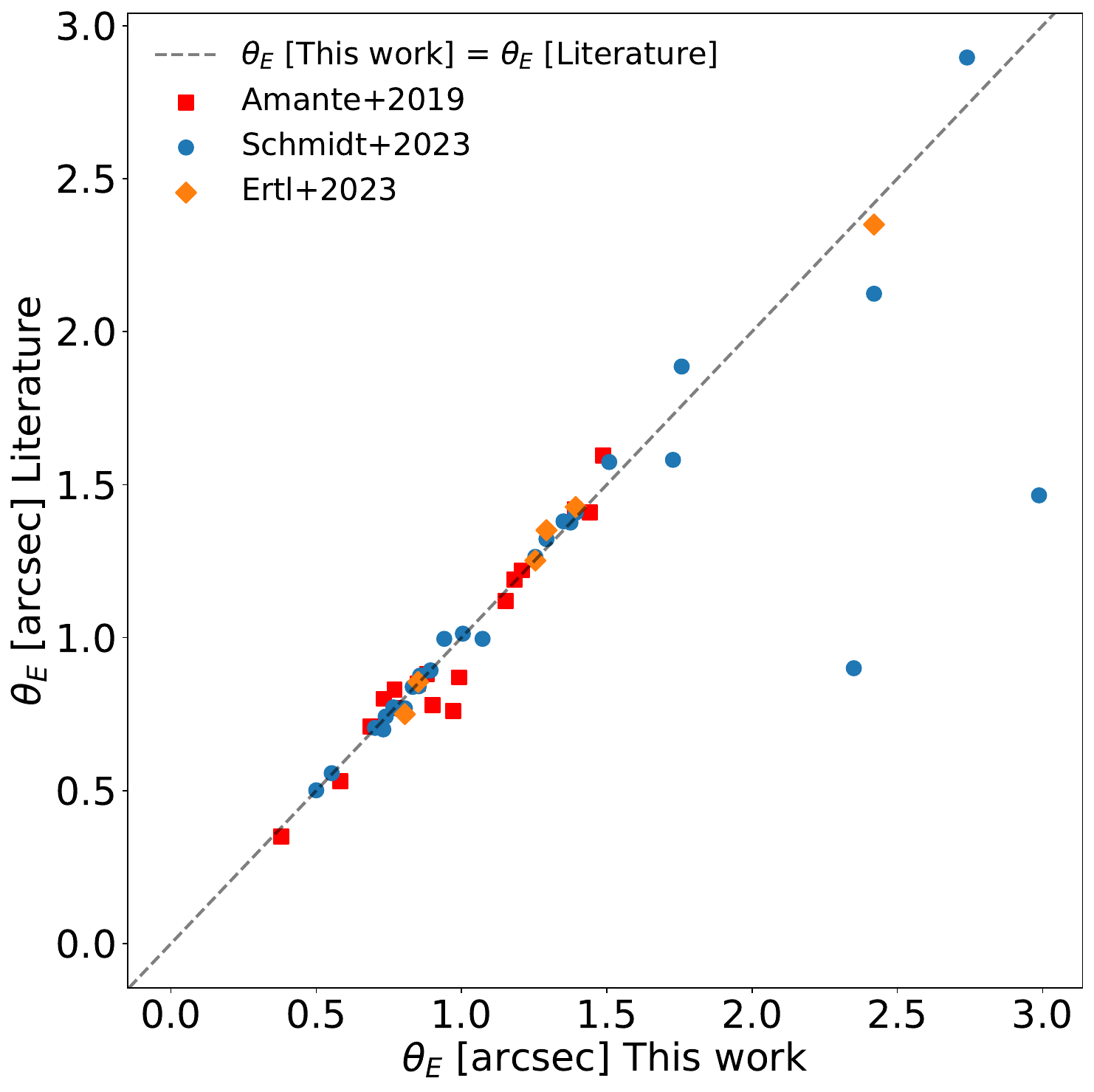}
    \caption{Comparison $\theta_E$ in this work vs values in the literature for modeled lens quasars. The systems that are clearly far from a 1:1 relation are SDSS 1155+6346 and HS 0818+1227.}
    \label{fig: comparison of thetas}
\end{figure}

\subsection{Microlensing magnification maps \label{sec:maps}}
The Einstein radius defined in the previous section applies to a single, isolated microlens. However, microlensing of lensed quasar images is associated with a relatively high optical depth of the microlenses, and cooperative effects between them are expected. To estimate the probability and time-scales of microlensing in this scenario, we are going to use microlensing magnification maps. The Probability Density Functions (PDF) of microlensing magnification obtained from these maps provide information about the probability of brightness anomalies with respect to the intrinsic flux-ratios between images.

The analysis of magnification maps is a vital component of microlensing studies. In recent years, developments in this field have advanced significantly, particularly through the incorporation of GPU-based methods \citep[e.g.][]{2025Weisenbach} and large public databases such as GERLUMPH \citep{2014Vernardos,2015Vernardos}, which contain thousands of high-resolution magnification maps, as well as through updates to classical techniques, such as the FMM–IPM \cite{2022JimenezVicente}, an improved version of the classical inverse polygon mapping method \citep{2006Mediavilla,2011Mediavilla}. 
In this study, we employ the FMM–IPM, which has been used extensively in microlensing analyses \citep[e.g.][]{2025JimenezVicente,2024Fian,2023EstebanGutierrez}, and whose magnification maps can be queried online \footnote{see \url{https://gloton.ugr.es/microlensing/}}. 
\begin{figure}[h]
    \centering
    \includegraphics[width=1\linewidth]{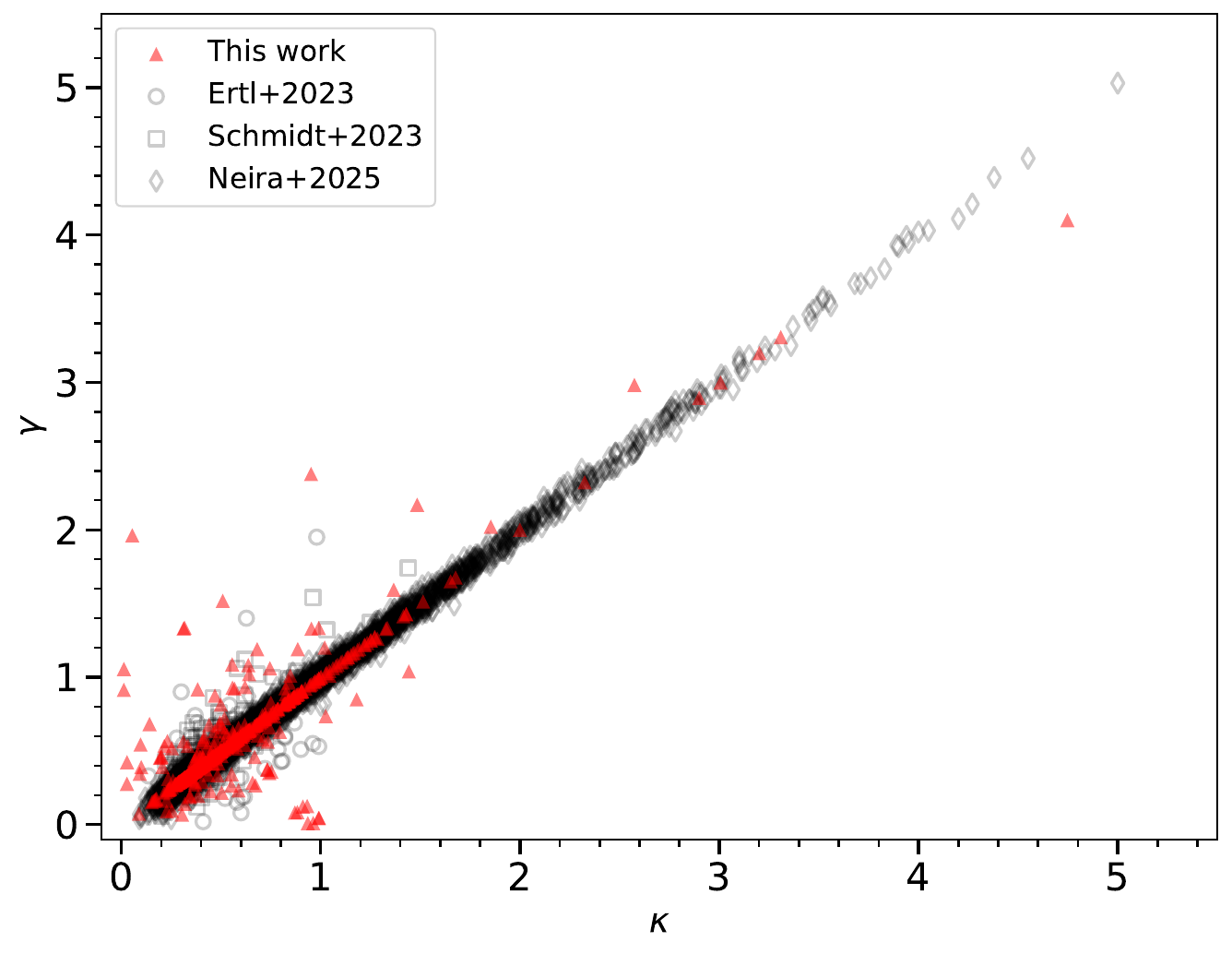}
    \caption{Values of $\kappa$ and $\gamma$ for our sample (red filled diamonds), compared with literature samples: unfilled black diamonds \cite{2025Neira}, unfilled circles \cite{2023Ertl}, and unfilled black squares \cite{2023Schmidt}.}
    \label{fig:kappa_gamma_dist}
\end{figure}

The parameters used to generate the magnification maps include the convergence, $\kappa$, and shear, $\gamma$, which are obtained from the lens modeling (their distributions are shown in Figure~\ref{fig:kappa_gamma_dist}); the fraction of the total convergence in microlenses, defined as $\alpha=\kappa_*/\kappa_{\rm tot}$; and the microlens Einstein radius (see Equation~\ref{eq:Re}). The magnification maps have a physical size of 60 Einstein radii, with pixel dimensions ranging from 400 (Q2237+030) to 3800 (WISE2329-1258), corresponding to an average scale of approximately 1.1 light-days per pixel. This resolution is sufficient to avoid introducing additional biases in our results, as demonstrated by \cite{2013Vernardos}.

Magnification maps are generated for two values of $\alpha$. We select $\alpha=0.1$ and $\alpha=0.2$, taking into consideration a low stellar mass fraction that has been reported in some microlensing studies, with values around a few percent to 10\% \citep{2009Mediavilla}, consistent with lens galaxies being largely dominated by dark matter at the radius of the quasar images. Then, in more recent analyses that include finite-source effects and larger samples, higher stellar mass fractions are indicated, converging on $\alpha \sim 0.2$ as a representative value near the Einstein radius \citep{2015JimenezVicente}. This result is also in line with strong-lensing and dynamical studies of early-type lens galaxies, which suggest that stars contribute roughly 10–30\% of the surface mass density at a few effective radii \citep[e.g.,][]{2024Vernardos}. Thus, by adopting $\alpha = 0.1$ and $0.2$, we are bracketing the most plausible range of stellar mass fractions inferred for lensing galaxies, while also ensuring comparability with previous microlensing studies that commonly use these values as benchmarks. An example of such a magnification map is shown in Figure~\ref{fig:mapa_magnification}.

To account for the magnification of a finite-size source, maps were convolved with a Gaussian source profile of standard deviation $r_s$, $I(r)\propto \exp{(-r^2/2r_s^2)}$. As shown by \cite{2005Mortonson}, the specific shape of the brightness profile does not strongly affect the statistical magnification properties, but just its half-light radius, which for a Gaussian profile is $R_{1/2}=1.18r_s$ and $R_{1/2}=2.44R_S$ (i.e., $r_s=2.07R_S$).

The $r_s$ values for each system are inferred from thin-disk estimates of $R_S$, rescaled to the median size derived from microlensing observations (see Section~\ref{sec:timescales}). 

For each system image, magnification maps are generated using the corresponding $\kappa$ and $\gamma$ values. The physical size of the maps varies among systems in order to preserve a fixed pixel resolution in light-days. We consider two values of the microlens mass fraction, $\alpha = 0.1$ and $0.2$, and three source-size scalings, $0.3\times r_s$, $1\times r_s$, and $2\times r_s$. For each set of ($\alpha$, $r_s$), this yields 520 magnification maps for the full sample consisting of 149 doubles, 53 quads, and 2 five-image systems. This gives a total of 3120 convolved maps for all the pairs ($\alpha$, $r_s$). These maps are then used to study the effects of $\alpha$ and $r_s$ on the magnification probability distributions and, combined with observational microlensing constraints, to evaluate the likelihood of different parameter combinations (Section \ref{sec:MMS}).

\begin{figure}
     \centering
     \includegraphics[width=0.5\textwidth]{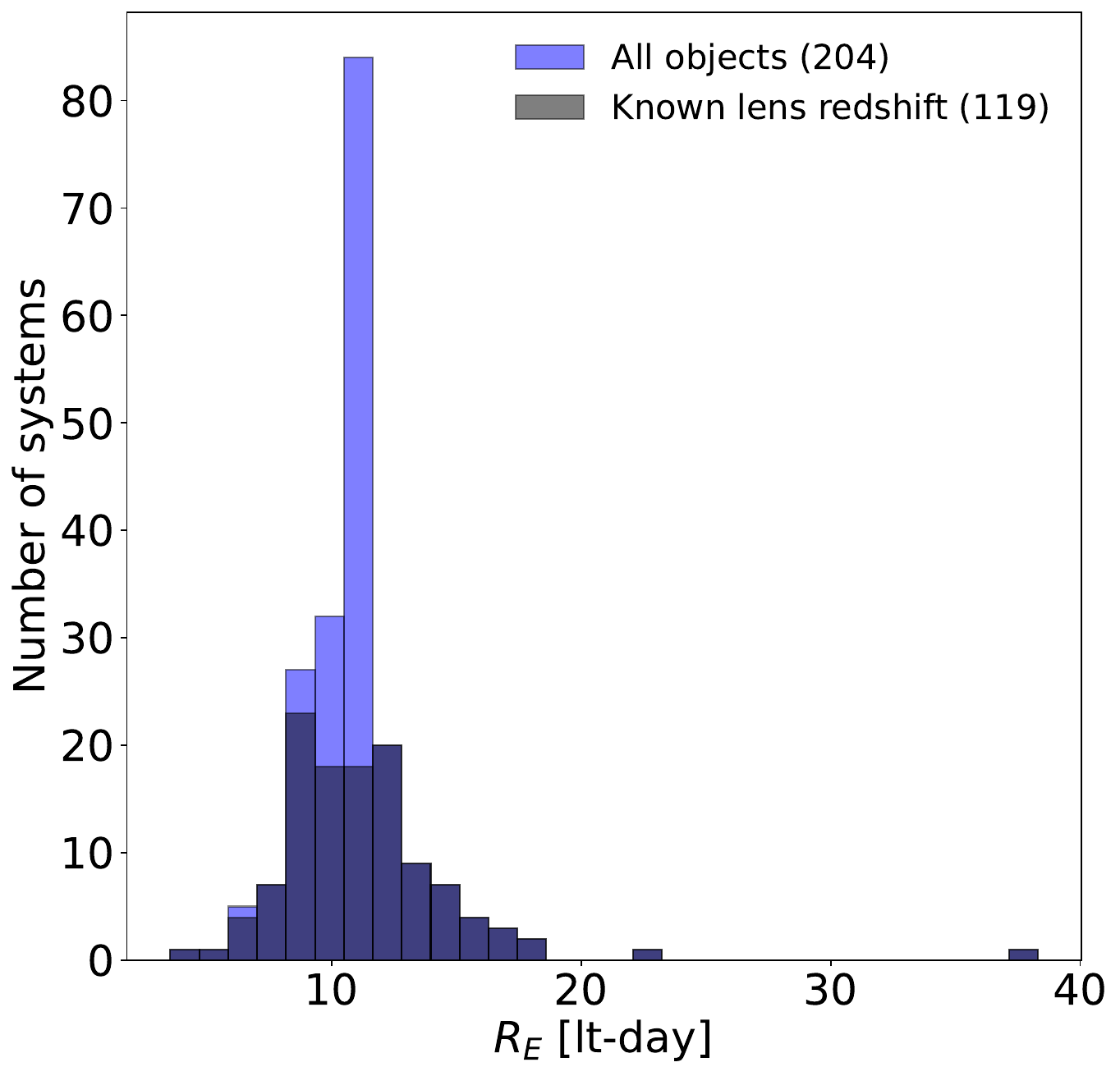}
     \caption{Histogram of the Einstein radii at the source plane for the systems, in blue the distribution of all the objects, and grey the distribution for the systems with the redshift of the lens known}
     \label{fig: re}
 \end{figure}
\section{Results}
\subsection{Homogeneous estimate of Einstein radius}
The homogeneous modeling of gravitational lensing systems provides an estimate of the parameters that define our model. The Einstein radius is the most relevant for our study, given its proportionality with the lens galaxy mass. To ascertain the robustness of our estimations, we will compare them with those obtained in the literature and with other codes.

\subsubsection{Comparison with previous studies not based on automatic modeling}

For comparison with the literature, we will use the sample of lensed quasars selected by \citealt{2020Amante}, which compiles the $\theta_E$ values individually obtained for each system (see Figure~\ref{fig: comparison of thetas}). The agreement is, in general, excellent, except for two systems. One of them, SDSS1155+6346 \citep{2004Pindor}, is a double lens quasar, and the model selected in the literature was chosen from \cite{2014Rojas}. In their work, they selected a SIS+$\gamma$ model while our model is an SIE for this system. The discrepancy may be related to the difference in mass distribution selection. The other case is HS0818+1227 \citep{2000Hagen}. This system is a double and the model selected, as cited in \cite{2011Leier}, is affected by a close external galaxy, which impacts the estimation of the Einstein radius. 

\subsubsection{Comparison with previous automatic modeling studies}
\label{ss: Comparison with previous automatic modeling studies}
The newest discovered quadruple systems have been automatically modeled using  HST high-resolution images. 
\cite{2023Schmidt} shows the results for thirty quadruply lensed quasars obtained with  \texttt{LENSTRONOMY}. We modeled 29 of the 30. The missing model system corresponds to J1721+8842, the first confirmed ZigZag lens \citep{2025Dux}, consisting of two lens galaxies at different redshifts. For the remaining 29, we obtain a good agreement for 27, with a difference in the $R_E<0.01$~arcsec (see Figure~\ref{fig: comparison of thetas}). The other two systems are J0343-2828 (Lemon in prep, which has a fifth image) and 2M1310-1714 (\cite{2018Lucey}, which also possesses a fifth image). We also compare the $\kappa$ and $\gamma$ obtained in our models with those obtained in \cite{2023Schmidt} we find $\Delta\kappa_{\mathrm{med}} \simeq 0.01$ and $\Delta\gamma_{\mathrm{med}} \simeq -0.01$.  Given the corresponding scatters, $\sigma_{\kappa,\mathrm{mad}} = 0.09$ and $\sigma_{\gamma,\mathrm{mad}} = 0.09$, these offsets correspond to $0.16\,\sigma$ in $\kappa$ and $0.06\,\sigma$ in $\gamma$, indicating good agreement with previous results.

On the other hand, \cite{2023Ertl} modeled 9 of the 30 systems using \texttt{GLEE}. Figure \ref{fig: comparison of thetas} demonstrates that eight of the nine systems (we also remove J1721+8842) analyzed in that study can be reproduced with excellent consistency (within 2~$\sigma$ for seven cases).

\subsection{Microlensing time scales \label{sec:timescales}}

Our sample can be divided into two main populations: one for systems with known redshifts (119) and the other with unknown redshifts for the lens (85). As previously commented, the microlensing time scale depends on the Einstein radius of the microlens, the source size, and the effective velocity, all measured at the source plane (see Section~\ref{sec:micro_time} for more detail). In Figure \ref{fig: re}, we present the histogram of the Einstein radii distribution for the two samples, which have medians of $10.900 \pm 0.138 \rm \,lt-day$ for the known redshift sample and $10.790 \pm 0.070 \rm\,lt-day$ for the entire sample assuming $z_L=0.5$ for the systems with unknown redshift. The 1~$\sigma$ uncertainty was inferred from the Monte-Carlo sampling of the results.
 
\begin{figure}[h]
    \centering
    \includegraphics[width=1\linewidth]{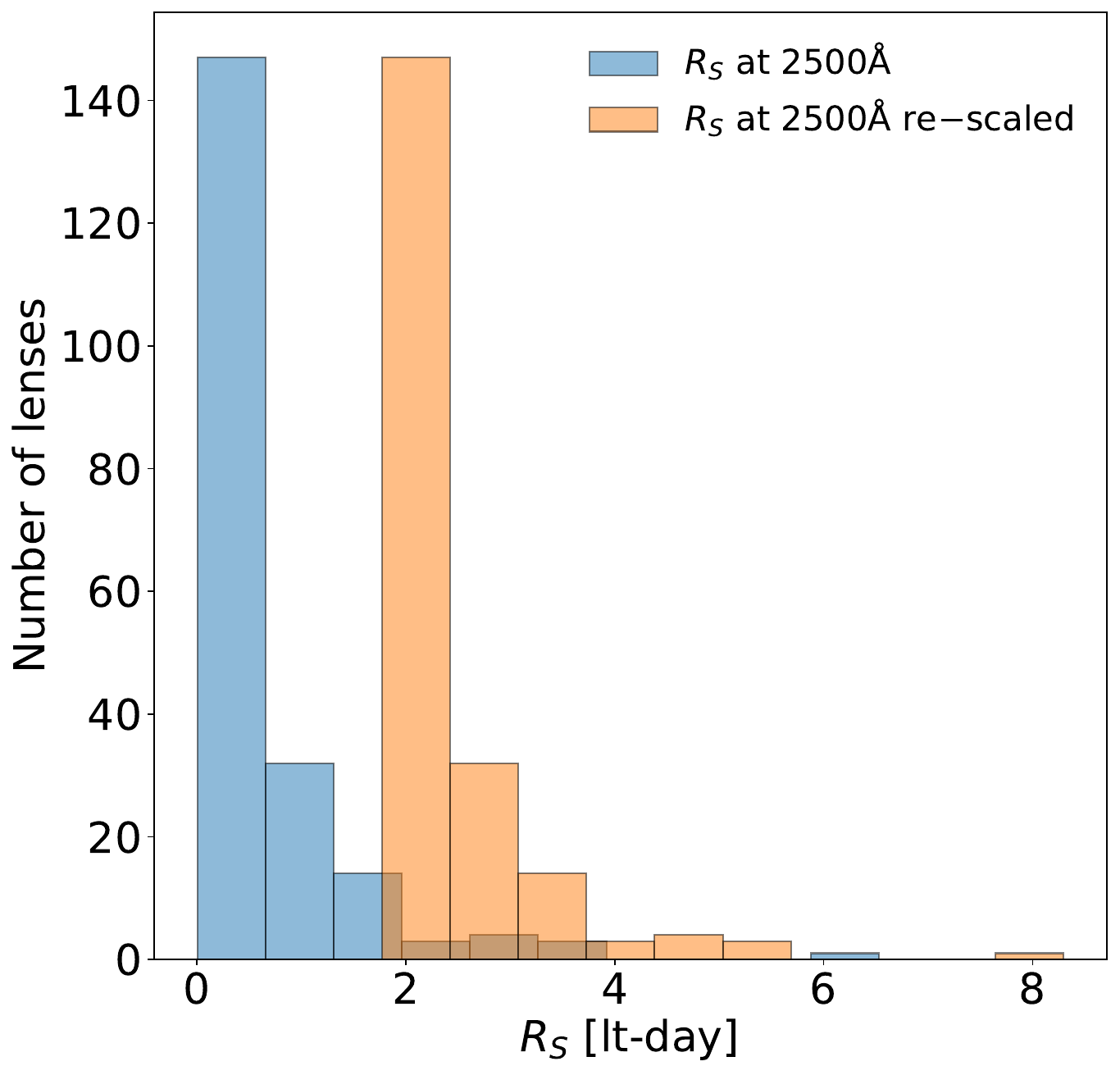}
    \caption{Histogram of the S\&S based sizes of the source, $R_S$, before and after applying microlensing size correction (204 systems).}
    \label{fig: rs}
\end{figure}

 In the case of the radius of the accretion disk ($R_{S}$), it is known that the size obtained by S\&S is not in agreement with observational estimates (obtained from reverberation mapping and/or microlensing) and usually underestimates the size of the accretion disk \citep{2010Morgan,2012JimenezVictente,2018Fausnaugh,2022Jha}. To account for this discrepancy, we first calculate the median and standard deviation of the microlensing-based values obtained by \cite{2020Cornachione} and \cite{2012JimenezVictente}. Then we re-scale all the values to a rest frame of $\lambda_{rest} = 2500 \r{A}$. Using Monte-Carlo sampling to estimate the 1~$\sigma$ uncertainties, we obtain a representative experimental median value of $R_S^{exp}= 1.77\pm 0.41$ $\rm\,lt-day$.  Then we use the ratio between this quantity and the median of our theoretical estimates of $R_{S}$, based on Equation \ref{eq:Rs},  $R_S^{theor}=0.23 \pm 0.003$ $\rm\,lt-day$, to obtain microlensing corrected $R_{S}$.  The resulting $R_{S}$ are shown in Figure~\ref{fig: rs}, with a median of  $2.23 \pm 0.06$ $\rm lt-day$ at 2500 \AA \,after re-scaling.

\begin{figure}[h]
    \centering
    \includegraphics[width=1\linewidth]{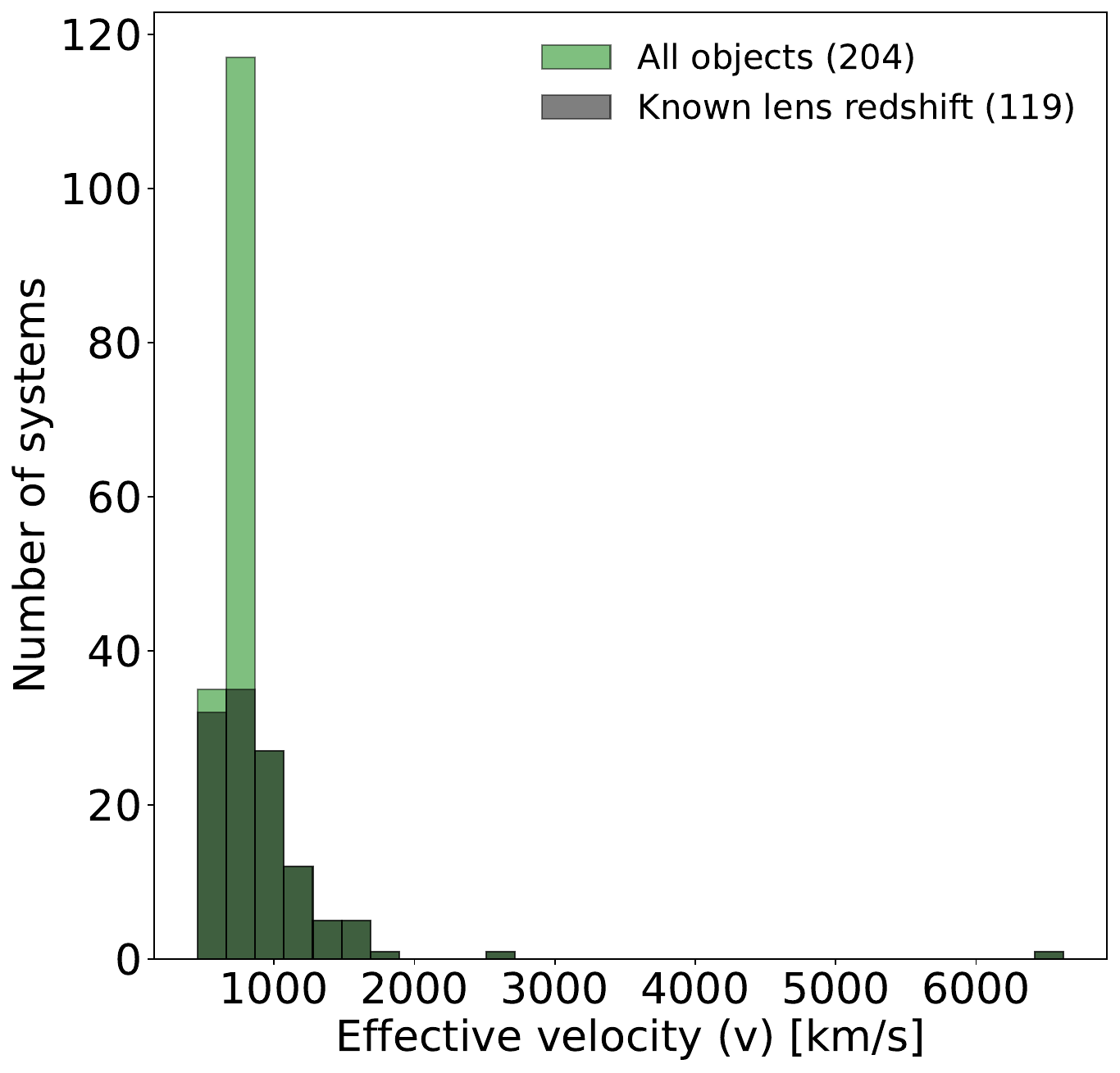}
    \caption{Histogram of the distribution of the effective velocities (Equation~\ref{eq:v_main}).}
    \label{fig:v}
\end{figure}

The last ingredient to calculate the time-scales is the effective velocity of the source (see Section~\ref{sec:micro_time}); we present our results in Figure \ref{fig:v}. The median value of the histogram is $ 807 \pm 17 $ $\rm km/s$ for the systems with known redshifts, and 778 $\pm$ 10 $\rm km/s$ for the whole sample.

\begin{figure}
    \centering
    \includegraphics[width=1\linewidth]{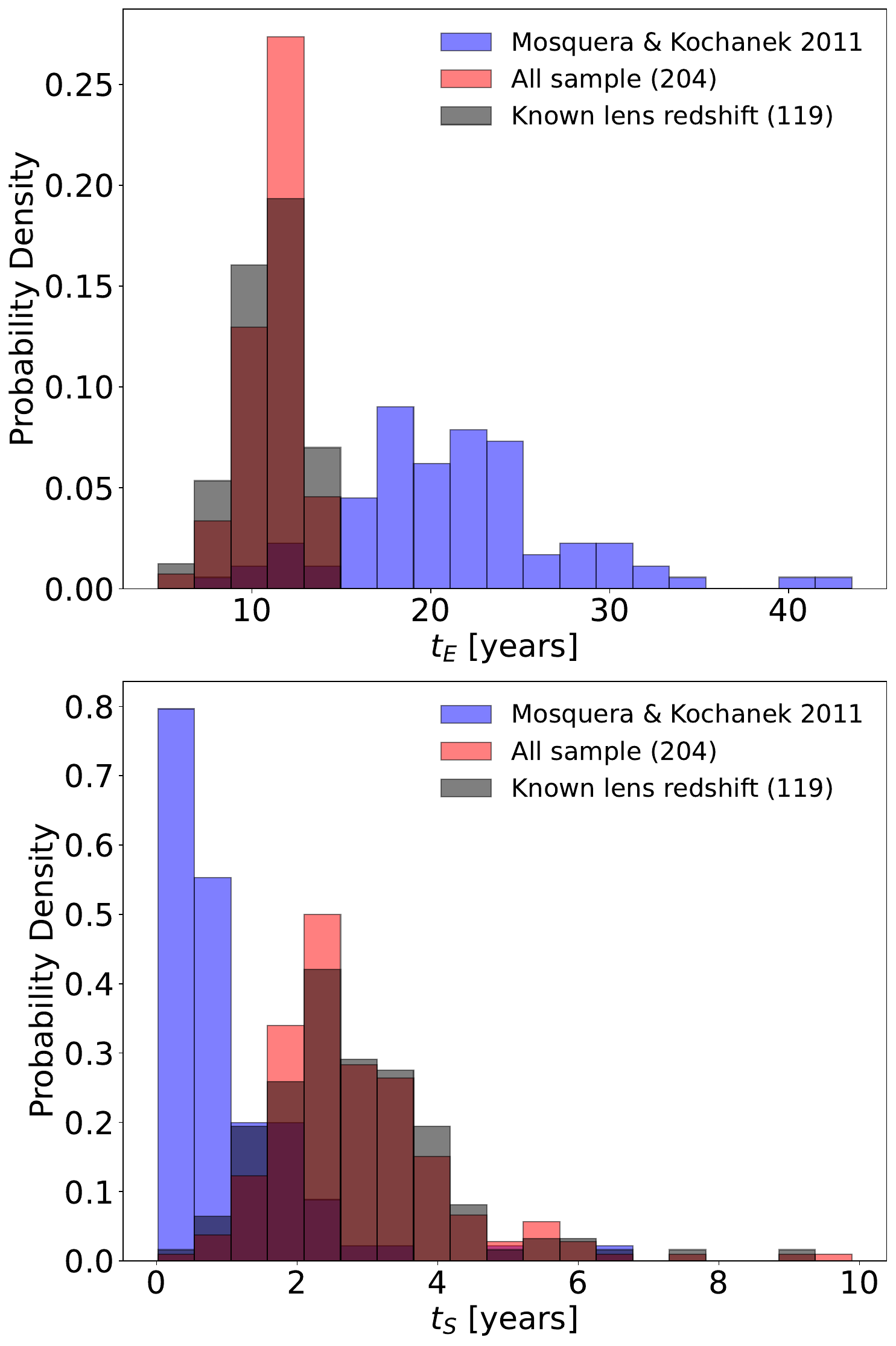}
    \caption{Top: Distribution of Einstein time scales.
    Bottom: Distribution of source time scales.}
    \label{fig:time_scales}
\end{figure}

Finally, to compare with \citetalias{2011Mosquera} results, we first need to rescale our median-corrected source sizes to the wavelength $8140{\rm \AA}/(1+z_{S})$, which is the original wavelength that \citetalias{2011Mosquera} use to present their results. Our values are in Figure~\ref{fig:time_scales}. From these histograms, we can derive, with 1~$\sigma$ uncertainties, that the mean Einstein ring crossing time is $11.29 \pm 0.05$ years for the entire sample and $10.82 \pm 0.20$ years for the systems with known lens redshifts, whereas \citetalias{2011Mosquera} obtained a value of $20.6$ years. The difference between the two estimates is mainly related to using the updated value for the peculiar velocity at zero redshift by \cite{2016Hess}. Notice that \cite{2024Mediavilla}, which uses a similar approach to estimate the peculiar velocities, obtains an average Einstein radius of 9.4 years, in good agreement with our mean value. On the other hand, for the source crossing time scale, $t_{S}$, we obtain a mean value of $2.59 \pm 0.07\rm $  and  $2.59 \pm 0.09 \rm $ years for known redshift lens systems and all the sample, respectively, in contrast with the $0.61 $ years calculated by \citetalias{2011Mosquera}. In this case, the discrepancy arises from both, the re-scaling of the theoretical sizes and the updated effective velocity.

\begin{table*}
\caption{Mean microlensing statistics for different stellar mass fractions and source-size scaling factors.}
\label{tab:stats}
\centering
\renewcommand{\arraystretch}{1.2}
\begin{tabular}{cccccc}
\hline
$\alpha$ & $\times\,\{r_{s}\}$ & Mean $|\Delta \mathrm{mag}|$ & $\hat{\chi}^2$ & Mean prob. $\Delta \mathrm{mag} < -0.32$ & Mean prob. $\Delta \mathrm{mag} > 0.32$ \\
\hline
0.1 & 0.3 & 0.227 & 9.362  & 12.193 & 20.043 \\
0.2 & 0.3 & 0.320 & 8.497  & 15.162 & 29.586 \\
0.1 & 1.0 & 0.153 & 9.203  & 6.309  & 17.815 \\
0.2 & 1.0 & 0.217 & 6.430  & 8.646  & 26.310 \\
0.1 & 2.0 & 0.096 & 16.316 & 1.923  & 15.984 \\
0.2 & 2.0 & 0.158 & 8.650  & 3.188  & 23.834 \\
\hline
\end{tabular}
\tablefoot{The table lists the mean values of $|\Delta \mathrm{mag}|$, the probability of $\Delta \mathrm{mag} < -0.32$, and the probability of $\Delta \mathrm{mag} > 0.32$ for different values of $\alpha$ and source-size scaling factors $\times\,\{r_s\}$. The samples were normalized so that bin counts sum to 44. The quantity $\hat{\chi}^2$ was calculated using Eq.~\eqref{eq:chia}.
}
\end{table*}

\begin{figure*}[h]
    \centering
    \includegraphics[width=1\linewidth]{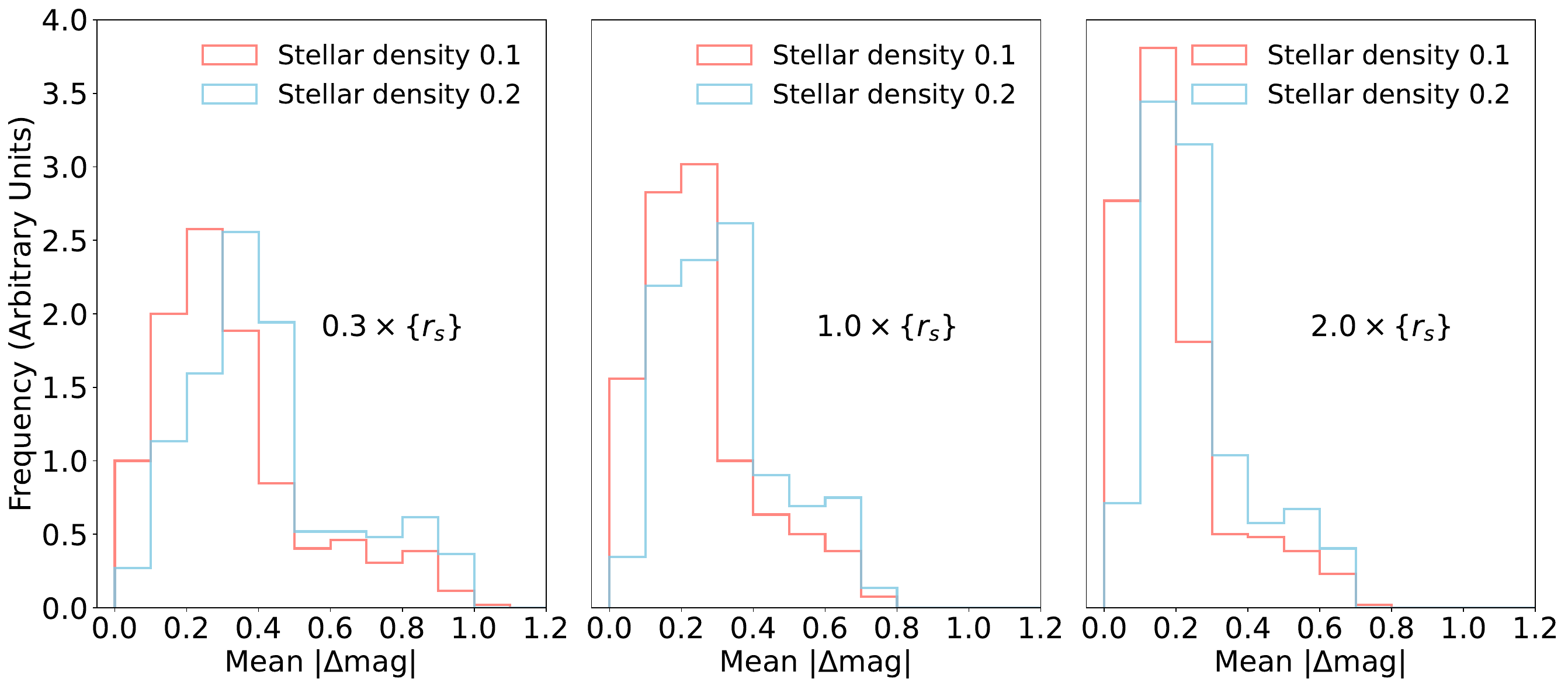}
    \caption{Probability distribution of unsigned microlensing magnification for three different scaling factor values for 0.3 $\times$ \{$r_s$\}, 1.0 $\times$ \{$r_s$\}, and 2.0 $\times$ \{$r_s$\}, with two different stellar densities ($\alpha$).}
    \label{fig:microlensing_compare_prenorm}
\end{figure*}

\subsection{Microlensing magnification statistics}
 \label{sec:MMS}
\begin{figure*}
    \centering
    \includegraphics[width=1\linewidth, height=9cm]{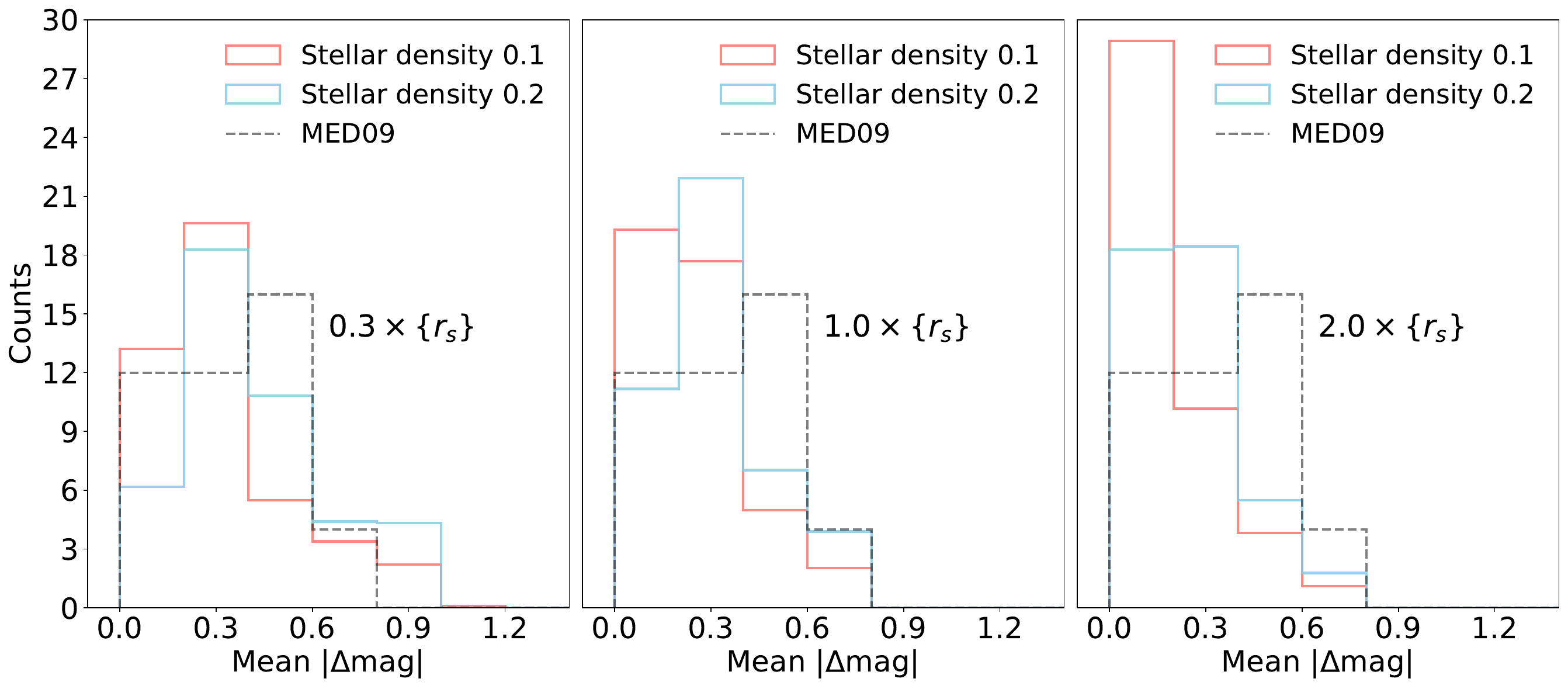}
    \caption{Histogram of unsigned microlensing magnification for three different scaling factors values for $r_{s}$, with two different stellar densities ($\alpha$), comparing the results from \cite{2024Mediavilla} and this work with all histograms normalized to the same total count.}
    \label{fig:microlensing_compare}
\end{figure*}

 To explore the possible impact of the source size, $r_s$ ($r_s=2.07R_S$), and of the fraction of mass in microlenses, $\alpha$, on the microlensing magnification statistics, we have calculated (for each image of the systems in the sample) magnification maps for two different values of the fraction of mass in microlenses (see Section~\ref{sec:maps}). These maps are convolved with Gaussian sources of three different sizes, obtaining $2 \times 3$ different magnification map models $(\alpha,r_s)$  and their corresponding magnification histograms.  Then, for each of the six individual PDFs obtained for each image, we have calculated the mean of the unsigned microlensing magnification values  \footnote{As discussed in Section 2, the empirical histogram of microlensing magnifications from \cite{2009Mediavilla,2024Mediavilla}, which we use to compare predictions with observations, is based on single-epoch measurements that do not determine the sign and are derived from image pairs, whereas our values are derived from individual quasar images.} . In Figure \ref{fig:microlensing_compare_prenorm}, we present the histograms of these means (520 maps each ($\alpha$, $r_s$) pair), which give us the probability of obtaining an (unsigned) value of microlensing magnification if we randomly select one image from the sample. The mean values of these histograms are listed in Table~\ref{tab:stats}. As it can be inferred from Figure \ref{fig:microlensing_compare_prenorm}, an important difference between the six histograms is the significance of the tail, which is basically ruled by the size of the source (the smaller the size the heavier the tail). This difference may be relevant because there is no evidence of a fat tail in the histograms of observed microlensing magnification \citep[see histogram MED09 in Figure 1 of][]{2024Mediavilla}. To explore this question in detail, we plot the six histograms in Figure \ref{fig:microlensing_compare} with a bin size matching that of \cite{2024Mediavilla} this is done for all the images in our sets, for each ($\alpha$,$r_{s}$) pair.  According to this figure,  larger values of the fraction of mass in microlenses and small sizes tend to predict larger magnifications, which were not observed in the MED09 sample. In contrast, small fractions of mass in microlenses and large sizes overpredict low magnifications. 

To quantitatively compare the different histograms with MED09, we use a 
 $\hat\chi^2$ statistics that takes into account the contribution to the sample variance of both, the modeled and the observed histograms,
\begin{align}
  \hat{\chi}_{}^2(\alpha,r_s)
    &= \sum_{i} \frac{(O_i - H_i)^2}{O_i + H_i}.
      \label{eq:chia} 
\end{align}
\noindent
 Here, $O_i$ and $H_i$ correspond to each bin from the observed histogram, and the modeled one respectively from \cite{{2024Mediavilla}}
\begin{figure}
    \centering
    \includegraphics[width=1.\linewidth]{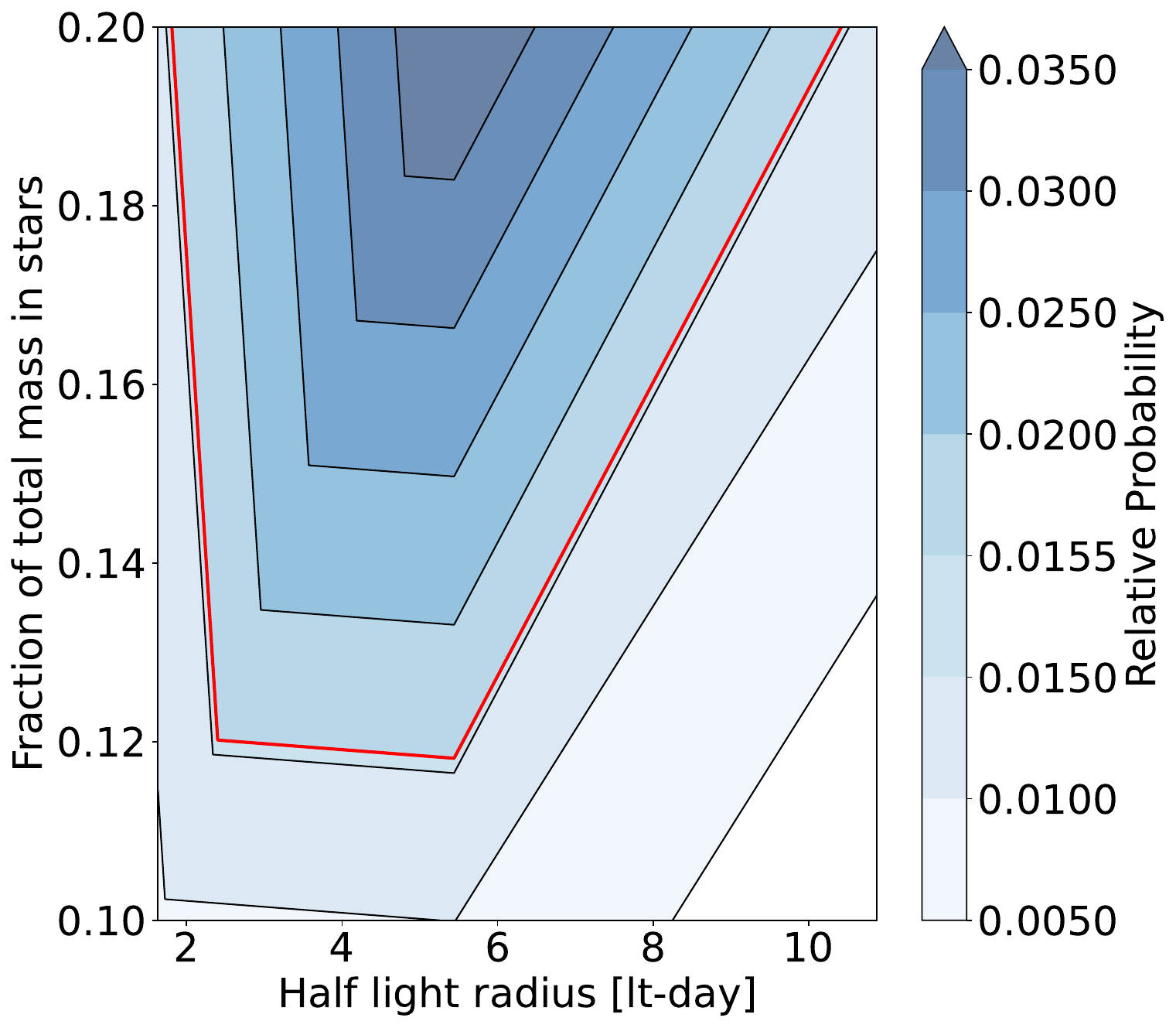}
    \caption{Relative probability distribution ($\propto e^{-\chi^2/2}$) in the (half light radius [light days], fraction of total mass in stars) parameter space. The red contour marks the 68\% confidence region, enclosing the highest-probability values.}
    \label{fig:probs_cont}
\end{figure}

 The $\hat \chi^2(\alpha,r_s)$ values obtained are listed in Table~\ref{tab:stats} and the relative likelihoods of the models, calculated from $L(\alpha,r_s)\propto \exp{(-\hat\chi^2(\alpha,r_s)/2)}$, are represented in Figure \ref{fig:probs_cont}. There is a clear minimum in $\hat\chi^2(\alpha,r_s)$ (maximum probability in Figure \ref{fig:probs_cont}), which corresponds to the model $\alpha=0.2$ and set of sizes $1\times \{r_s\}$. In Figure \ref{fig:probs_cont} we also plot the contour enclosing a 68\% of the probability, which leaves outside the models corresponding to sizes $2\times \{r_s\}$. Notice that there is a slight covariance between size and fraction of mass in microlenses in the sense that large sizes can be compensated by large mass fractions and vice versa. Marginalizing on the mass fraction, $\alpha$, we can obtain an estimate of the expected values for the size, $(1.0\pm0.5)\times \{r_s\}$, which, taking into account our median value $R_s=2.23$ light-days, results in $r_s=2.07R_S=4.6\pm 2.3$ light-days and in an half-light radius of $R_{1/2}=2.44 R_S=5.4\pm 2.7$ light-days. Marginalizing on size, we can infer a lower limit for the fraction of mass in microlenses of $\alpha\ge 0.15 \ (0.12)$ in the 68\% (90\%) confidence interval.

\begin{figure*}
    \centering
    \includegraphics[width=1\linewidth]{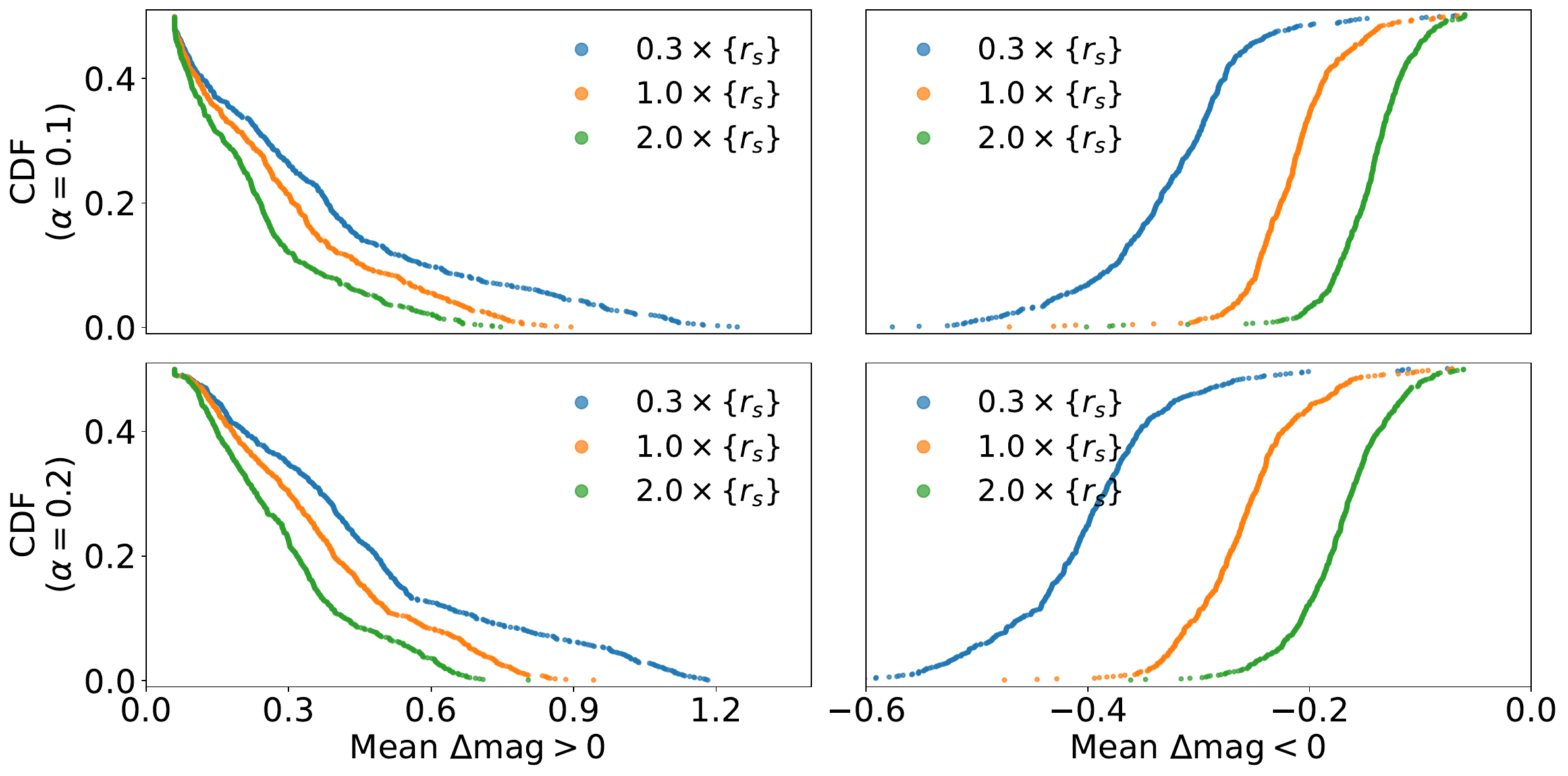}
    \caption{Figure: Cumulative distribution functions of magnitude changes during microlensing events. The left column shows the probability of observing a demagnification event with $\Delta \mathrm{mag} > 0$, while the right column shows a magnification event with $\Delta \mathrm{mag} < 0$. The top row corresponds to a stellar fraction of $\alpha = 0.1$, and the bottom row to $\alpha = 0.2$.}
    \label{fig:CDF}
\end{figure*}

 To go further with the statistical analysis of the histograms, we can focus on high magnifications, which are associated to interesting events like caustic crossings. In Figure \ref{fig:CDF} we show the cumulative distribution functions (CDFs) separating magnifications ($\Delta m < 0$) and demagnifications ($\Delta m > 0$). The most distinctive feature of these cumulative distributions is the size dependence of the probability of high magnifications. The integrated probability of events with $\Delta m < -0.40$ for sizes $0.3 \times \{r_s\}$ is a non-negligible 20\% (10\%) for $\alpha=0.2\ (0.1)$ while these high magnifications are very unlikely for larger sizes.

\begin{figure*}
    \centering
    \includegraphics[width=1.0\linewidth]{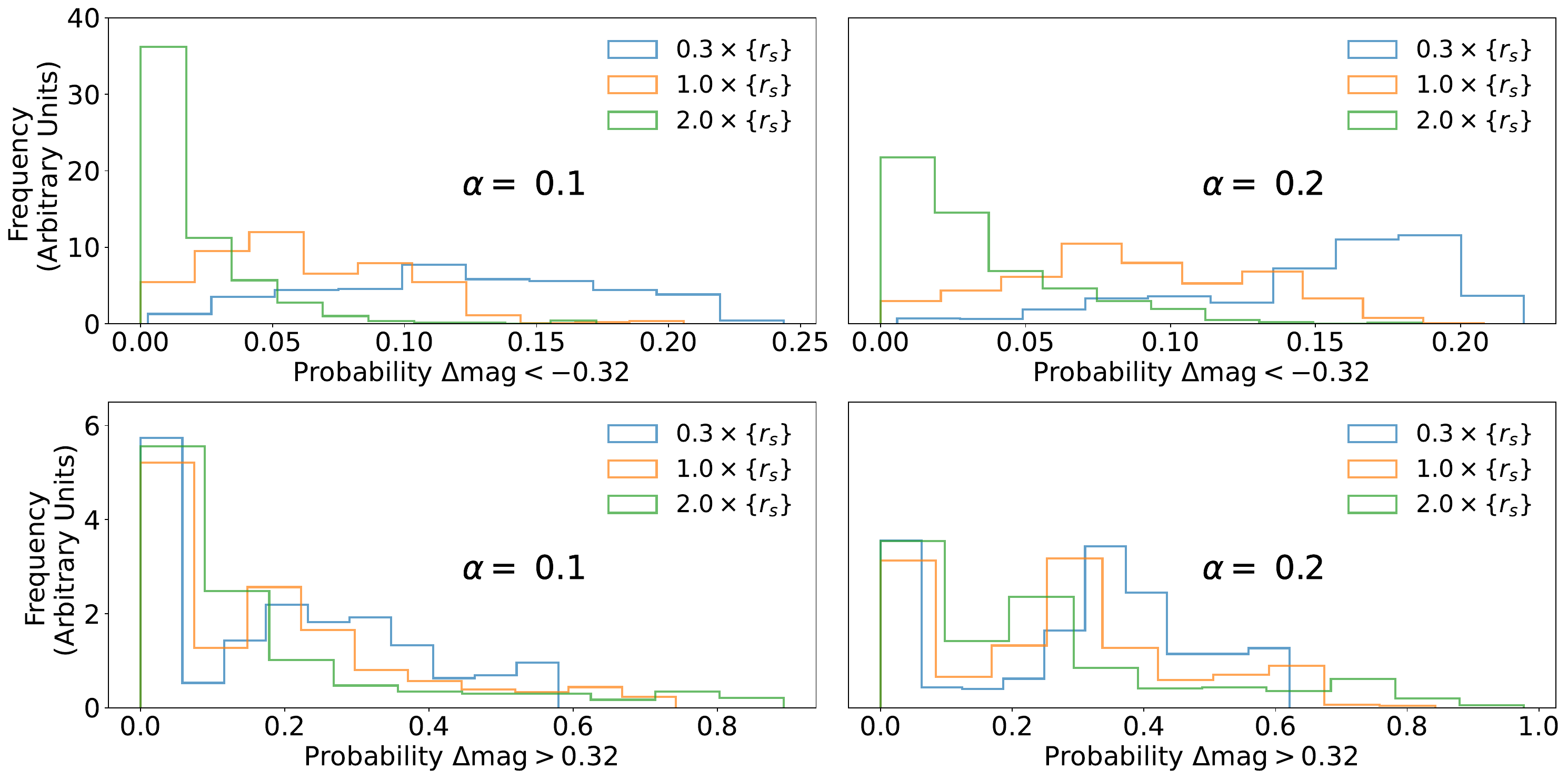}
    \caption{Top: histogram of the probabilities of obtaining a magnification less than $\Delta \mathrm{mag} = -0.32$ for different stellar densities and values of $r_s$.
    Bottom: density histogram of the probabilities of obtaining a demagnification greater than $\Delta \mathrm{mag} = 0.32$ for different stellar densities and values of $r_s$.}
    \label{fig:max_min_032}
\end{figure*}

\begin{figure*}[h]
    \centering
    \includegraphics[width=0.85\linewidth]{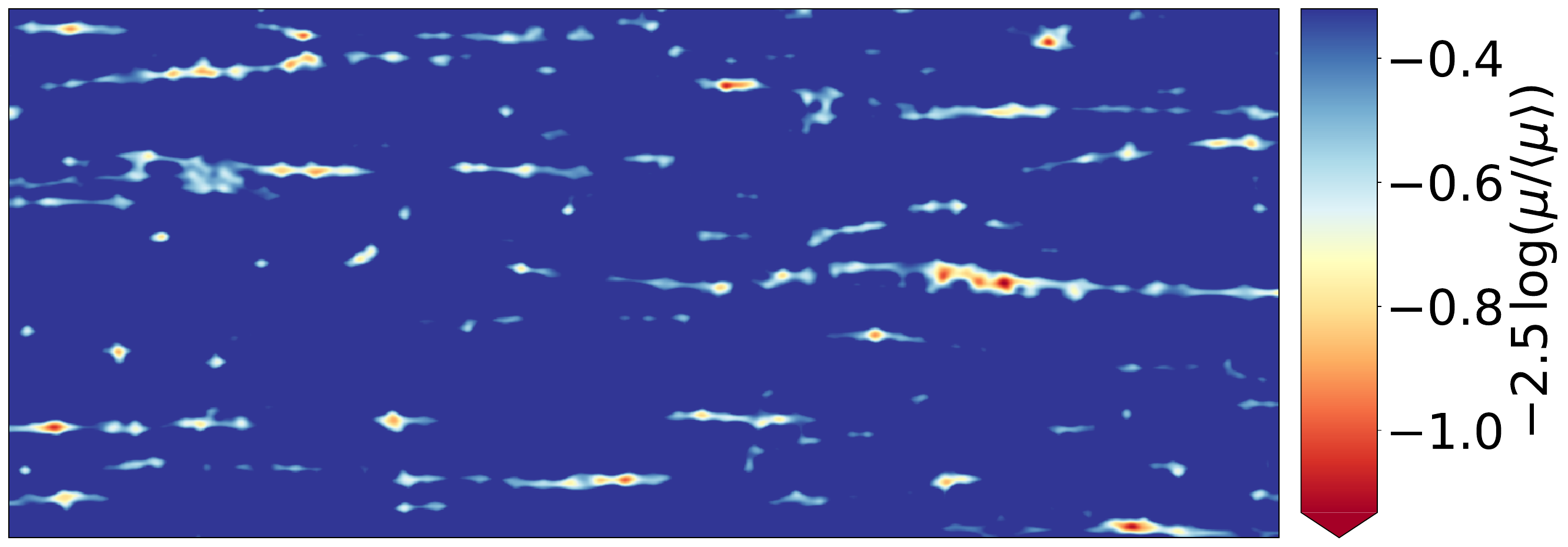}
    \caption{
    Same convolved microlensing magnification map as in 
    Fig.~\ref{fig:mapa_magnification} (Right~panel), 
    With zoom in between pixel 0 and 400 in the Y axis and full length (1200) in the X axis. 
    Here, the magnification map is masked for values over $-0.32$, 
    highlighting the regions of strongest magnification within the caustic network. 
    }
    \label{fig:microlensing_map_limit_n032}
\end{figure*}

\begin{figure*}
    \centering
    \includegraphics[width=1.\linewidth]{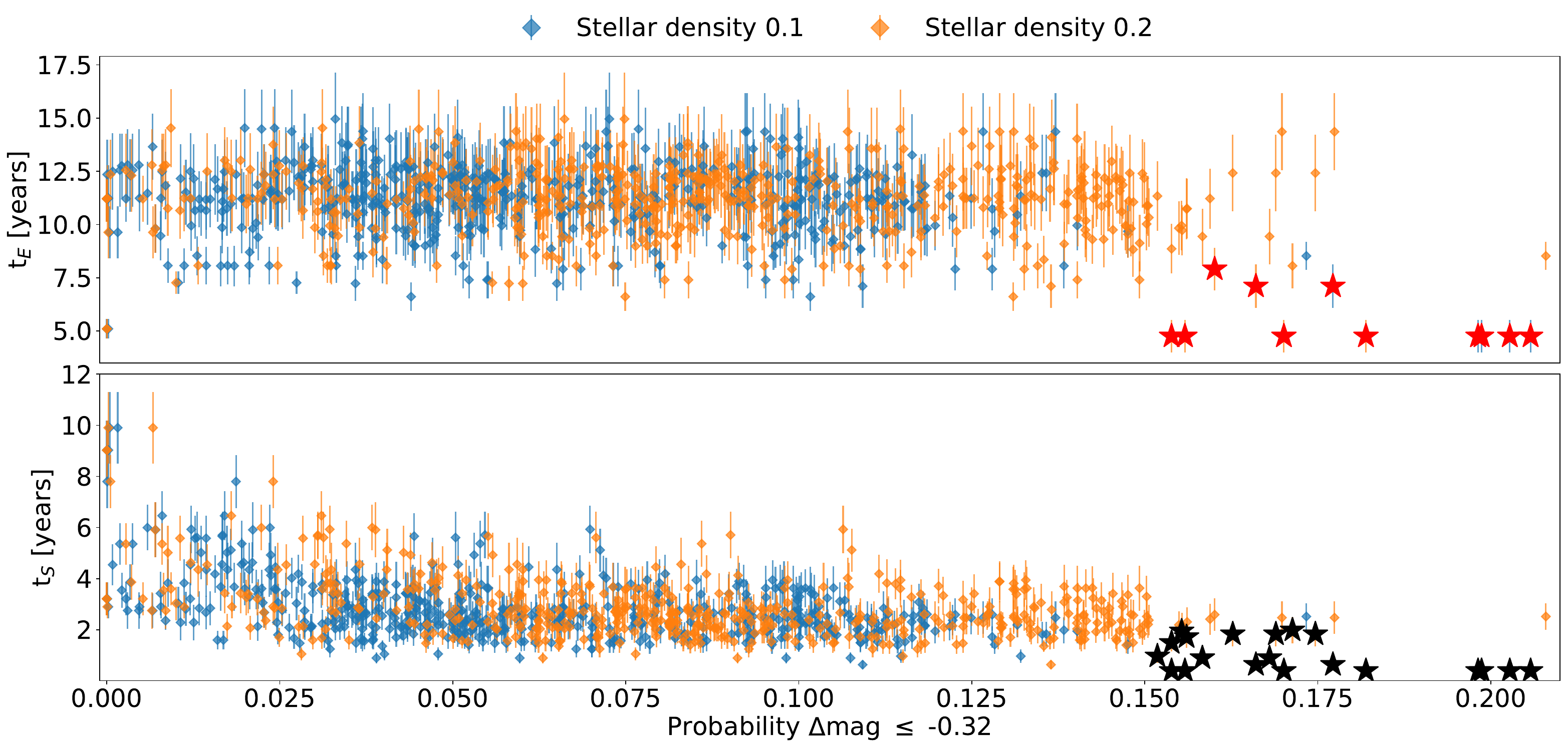}
    \caption{Top: probability of $\Delta \mathrm{mag} \leq -0.32$ versus Einstein radius crossing time (t$_{\rm E}$ [years]), with blue (orange) representing values obtained for a stellar density of 0.1 (0.2) and a 1.0 $\times\,\{r_{s}\}$, the red stars represent the object that shows the smaller Einstein radius crossing time and high magnification probability that correspond to DESJ2038-4008 and Q2237+030.
    Bottom: probability or $\Delta \mathrm{mag} \leq -0.32$ versus source crossing time (t$_{\rm S}$ [years]), with blue (orange) representing values obtained for a stellar density of 0.1 (0.2) and a R$_{S} \times$ 1, in black stars the objects that show the smallest source crossing time and hight probability that are CY2201-3201, WG2100-4452, SDSSJ1138+0314, SDSSJ0821+4542, J0343-2828, WG0214-2105 and Q2237+030.}
    \label{fig:above-0.32magvscrossing_times}
\end{figure*}

To estimate the probability of an HME, in Figure \ref{fig:max_min_032} we plot the probability of having a magnification\footnote{We select the umbral magnification $\Delta m = -0.32$, as it is the value at the Einstein ring of an isolated particle.} $\Delta m \le -0.32$ and a demagnification $\Delta m \ge 0.32$, respectively.  As can be seen in this Figure, the probabilities for magnification are dependent on the source size and the fraction of mass in microlenses. On the contrary, there is no significant dependence in the case of demagnification. The mean probabilities for the case ($\alpha=0.2, 1\times \{r_s\}$) are  9\% ($\Delta m \le -0.32$) and  26\% ($\Delta m \ge 0.32$). When microlensing is detected from a flux ratio, it is not possible, a priori, to separate magnification from demagnification and, consequently, the probability of observing microlensing magnification of amplitude $|\Delta m| \ge 0.32$ can be  significantly larger, even of a  35\%, but at most in one-third of cases the large magnification may eventually be related to a singular magnification event such as a caustic crossing.

\section{Discussion}
\subsection{Average size of the quasar source and properties of the microlenses population.}
 We find a good matching between the experimental microlensing data from MED09 and the histogram of microlensing magnifications corresponding to the case $\alpha=0.2$ and average 
half-light radius $R_{1/2}=$ 5.4 light-days modeled by us for the available sample of lenses. These values are in good agreement with previous results from microlensing and reverberation mapping studies. Specifically, from the comparison between histograms, we obtain an estimate of the source size, $R_{1/2}=5.4\pm 2.7$ light-days, which is only slightly affected by the degeneracy with the fraction of mass in microlenses that has been pointed out in other studies.  This value for the size is in good agreement with the average size estimate related to reverberation mapping observations  \citep[see the discussion in][]{2017Mediavilla}. This result is quite robust, as a significant reduction in size would result in a fat tail of the histogram above $\Delta m\ge -0.5$ not observed in single epoch microlensing measurements (MED09) or in other observations based on light curves \citep[see the histogram PEAKS in][]{2024Mediavilla}. In what respect to the mass fraction in microlenses, we estimate a lower limit, $\alpha\ge 0.15$, also in good agreement with previous studies \citep[see, e.g.,][]{2022EstebanGutierrez,2023EstebanGutierrez}. To establish more stringent limits, larger values for $\alpha$ should be considered, which implies a computational challenge. We defer this study for future work, although values of $\alpha$ above 0.3 are unexpected. 

Regarding the possible existence of a population of intermediate mass Black Holes (BHs) contributing to micro or millilensing, notice that by virtue of the mass-length degeneracy (see Section~\ref{sec:introduction}), for relatively large microlenses masses ($\gtrsim 10 M_\odot$) the quasars will behave as point sources \citep[see, e.g.,][]{2023EstebanGutierrez,2024Heydenreich} and, consequently, a fat tail of high magnifications should be present in the histogram of observations. The absence of this tail excludes the existence of a significant population of BHs above this mass.

Finally, the good matching found between the histogram of magnifications modeled from the full sample of lens systems available and the histogram of observed microlensing magnifications (inferred from just a subset of systems) suggests both: that the histogram MED09 is statistically representative of microlensing magnification properties of lensed quasar systems and that the procedure followed by us to infer the magnification statistics, in particular lens modeling to obtain $\kappa$ and $\gamma$ for each lensed image, is reliable.

A caveat can be made regarding the comparison with MED09. These data
correspond to differential microlensing between pairs of images, which
may introduce some broadening of the histogram relative to single-image
microlensing. However, the good agreement between MED09 and the
distribution of microlensing peaks from light curves \citep[based on single
images; see][]{2024Mediavilla} indicates that this broadening is
small compared to other factors, such as intrinsic variability or
observational uncertainties.

\subsection{Magnification statistics and flux-ratio anomalies.}

A crucial step in many studies related to multiple imaged quasars is 
the modeling of the lens, which needs to be very precise, particularly for the
prediction of gravitational time delays between the images,
which may be used to solve the current tension in the
determination of the Hubble constant. However, calculated models
very often present strong differences between their predicted
flux ratios and the observed ones. These anomalies are often attributed to microlensing. However, the reported model anomalies \citep[see Figure 1 in][]{2024Mediavilla} are very high with mean values above 0.5 magnitudes and fat tails reaching even 2.5 magnitudes. According to our statistical study (see Figure \ref{fig:microlensing_compare_prenorm}), so high magnifications are not predicted even for unexpectedly small source sizes. Thus, we confirm for the present sample the results of previous simulations \citep{2022EstebanGutierrez,2023Awad}.
We provide microlensing magnification histograms for each individual image, which can be used to estimate the potential impact of microlensing on flux ratio anomalies as a criterion for rejecting or accepting a lens model.

\subsection{Microlensing time scales and frequency of HME.}

One main motivation of the statistical study of lens systems is to know with what probability one lens system can be experimenting high microlensing magnification. In \citetalias{2011Mosquera}, the answer to this question was based on the comparison between the Einstein and source crossing times. They conclude that half of the lensed images will be active with the source lying in the caustic ridges. 

Using our microlensing magnification statistics, we can now quantitatively set both the threshold to define an HME and the probability of exceeding it. In Figure  \ref{fig:max_min_032}, we show the probabilities histogram of having microlensing magnification larger than -0.32, which is the threshold to be within the Einstein ring in the case of an isolated microlens. As shown in Figure \ref{fig:max_min_032}, the average probability is around 9\% (the average probability of demagnification, $\Delta m \gtrsim 0.32$, associated with smooth variability, is about 26\%). We can compare our high magnification probability with the work of \cite{2025Neira}, where they estimate the number of HME events with magnification larger than 0.3$~\Delta$mag  by year for simulated light curves, obtaining $\sim~$60 for 1250 lens quasar images, which is $4.9\%$ per yr, this difference can be associated to the differences in sample and methodology.
\begin{figure*}[h]
    \centering
    \includegraphics[width=1\linewidth]{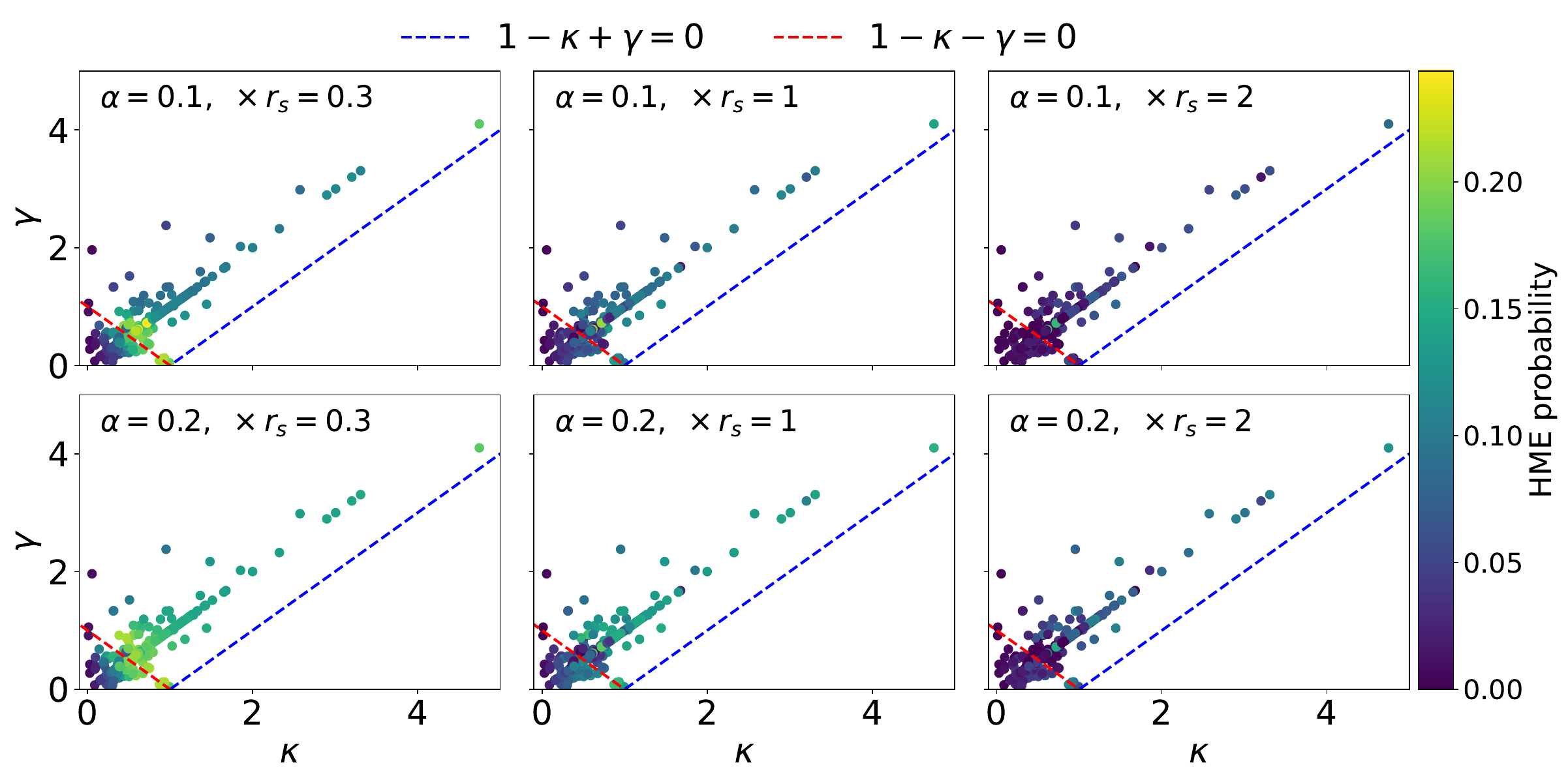}
    \caption{ Correlation between the values of $\kappa$ and $\gamma$ in our sample, with the color bar representing the HME probability. The panels are separated by $\alpha$ and by multiples of $r_s$.}
    \label{fig:HME_kappa_gamma}
\end{figure*}

It is also interesting to explore the possible correlation between the HME probability and the macro magnification of the image. With this aim, in Figure~\ref{fig:HME_kappa_gamma} we plot the HME probability (i.e. the probability that $\Delta \rm mag < -0.32$) versus $\kappa$ and $\gamma$ for each image in the sample. In this Figure we can observe (for the smallest size, $\times r_s=0.3$, considered in the calculations) a clear correlation of the HME probability with the high magnification region defined by the $1-\kappa-\gamma=0$  curve. This correlation disappears for larger values of the source size. The explanation of the correlation observed for $\times r_s=0.3$ is that a large macro-magnification of the image induce more structure in the microlensing magnification map, increasing the contrast between high magnification ridges and de-magnified valleys \cite[see, e.g.,][]{2021Weisenbach,2024Weisenbach,2025Neira} and, hence, the probability of HMEs.  However, this structure is washed out when convolved with a source of relatively large size.

On the other hand the origin of the discrepancy with the $\sim$50\% obtained by \citetalias{2011Mosquera} comparing $t_E$ and $t_S$, is that for optical depth close to 1, the real magnification map cannot be understood in terms of a single microdeflector because cooperative effects among microlenses give rise to the highly magnified caustic networks at expenses of other regions, which become demagnified, i.e., the scale of regions in the magnification map fulfilling $\Delta m = -0.32$ is smaller than one Einstein radius (see Section~\ref{sec:introduction}). 

In Figure \ref{fig:microlensing_map_limit_n032}, we show a magnification map where the low magnification pixels with $\Delta m > -0.32$ have been masked. Notice that the remaining regions appear to be the caustics and inner to caustic regions, and that they occupy just a fraction of the whole map. As discussed above, the apparent ‘caustics on top of caustics’ pattern reflects the densification and overlap of microcaustics for high-macro-magnified images (see Figure~\ref{fig:HME_kappa_gamma}), where the caustic network becomes connected and corrugated
\cite[see, e.g.,][]{2021Weisenbach,2024Weisenbach,2025Neira}. We have confirmed these results by inspecting magnification maps of 10 images uniformly selected according to the distribution of lensed images in Figure \ref{fig:above-0.32magvscrossing_times} (top).

In Figure~\ref{fig:above-0.32magvscrossing_times}, we show, for each one of the images of lens systems in our sample, the probability of microlensing magnification above -0.32 mag vs. crossing times. If we are interested in caustic scanning of the accretion disk or the innermost region of the BLR \citep{2024Fian}, we can select as candidates the objects with the highest probability and smaller source crossing time, for instance: CY2201-3201, WG2100-4452, SDSSJ1138+0314, SDSSJ0821+4542, J0343-2828, WG0214-2105, and Q2237+030.

If we are interested in estimating any pseudo-periodical property of microlensing \citep[like the caustic crossing frequency related to the peculiar velocity of the lens galaxy, see][]{2016Mediavilla}, we should, instead, select the systems with smaller Einstein radius crossing time, for instance: DESJ2038-4008 and Q2237+030.

\section{Conclusions}

With the aim of conducting a statistical study of some relevant properties of quasar microlensing, we collect data from all known lensed quasar systems and perform a comprehensive and homogeneous modeling of gravitational lensing systems to estimate key parameters, particularly the Einstein radius of the lensed systems, and the convergence ($\kappa$), and shear ($\gamma$), of the lensed quasar images. From $\kappa$ and $\gamma$ we obtain microlensing magnification maps for all the images using two values for stellar density (0.1 and 0.2). We combine data and calculations to estimate microlensing magnification probabilities and crossing times. The main results are the following:

1. The results of our lens fitting procedure demonstrate a high level of agreement with previous studies. Specifically, when compared with the sample of lensed quasars compiled by \citet{2020Amante}, we found that 88\% of the systems are consistent within our estimated uncertainties. Further comparisons with automatic modeling studies, such as those by \citet{2023Schmidt}, who used \texttt{LENSTRONOMY} on 30 quadruply lensed quasars, showed good concordance for 21 out of 29 common systems. Similarly, when compared with models from \texttt{GLEE}, our results agreed within 2~$\sigma$ for seven out of eight systems, reinforcing the reliability of our modeling approach.

 2. From the comparison of the observed microlensing magnification histogram with the magnification histogram of our sample of 204 lensed quasar systems modeled for different values of the fraction of mass in microlenses, $\alpha$, and the size of the quasar source, we obtain a Bayesian estimate of the average half-light radius $R_{1/2} = 5.4 \pm 2.7$ light-days and a lower limit $\alpha\ge 0.15$, in excellent agreement with previous studies based on microlensing and reverberation mapping.

3. According to the histogram of mean magnifications, it is unlikely that microlensing causes, on average, flux-ratio anomalies above $\sim$0.3 mag, and lens modeling leading to larger flux-ratio anomalies should be regarded with caution (in particular in the context of Hubble constant estimate from lensed quasar monitoring). The individual microlensing magnification PDFs that we provide and are available upon request for each image can be used to cross-check lens models by comparing the resulting flux-ratio anomalies with the predicted microlensing.

4. We found a median time scale $t_E = 11.29 \pm 0.05 \rm\,years$ for the Einstein radius crossing time and $t_S = 2.59 \pm 0.07 \rm\,years$ for the source size crossing time after accounting for uncertainties through Monte Carlo simulations. These values differ from those reported by \citetalias{2011Mosquera}, who found median values of 20 years and 0.61 years, respectively. The discrepancies can be mainly attributed to differences in the methodologies for calculating effective velocities and to our re-scaling of the theoretical accretion disk sizes to account for recent measurements based on reverberation mapping and microlensing.

5. Beyond the single particle-based analysis of the Einstein radius crossing time, the study of microlensing magnification maps and histograms allows us to obtain more detailed and quantitative information about the probability of observing some interesting phenomenology related to microlensing, like caustic crossing. Specifically, due to the formation of caustics networks, the size of the regions with magnification $\Delta m \le -0.32$ (corresponding to the region inner to the Einstein radius of one isolated particle) is significantly smaller than the Einstein radius. We find that these regions correspond to the caustic and (partly) to the inner to caustic locations and determine that these regions occupy, on average, 9\% of the magnification maps  (i.e., if you randomly pick an image from our sample, this is the rough probability for this image to be engaged in a past or future caustic crossing event). In agreement with previous studies \cite[see, e.g.,][]{2021Weisenbach,2024Weisenbach,2025Neira} we find a correlation between the macro-magnification of the image and the probability of HMEs for sources of relatively small size.

6. Combining the crossing times with the probabilities of a high magnification event, we determine the best candidates for dissecting the structure of accretion disks through caustic scanning and studying the statistics of pseudo-periodical magnification events (like caustic or zero crossings).

\section*{Data availability}
In the body of the paper, we presented Figures and results for the scales, effective velocity, source sizes, and magnification map statistics. We provide an online appendix at \url{https://github.com/felavila/microlensing-Maps-Post-Analysis-results}.

\begin{acknowledgements}
    We thank Jorge Jiménez Vicente (†) for his valuable help and insightful comments.
    FAV thanks the Doctorate Fellowship program FIB-UV of the Universidad de Valparaíso for funding, ESO Early-Career Visitor Programme and the ESO Science Support Discretionary Fund. This work was supported by ANID FONDECYT Regular grant number 1231418 (VM, FAV) and by the Centro de Astrofísica de Valparaíso (CIDI 21). EM acknowledges support from grants PID2020-118687GB-C31 and PID2020-118687GB-C33, financed by the Spanish Ministerio de Ciencia e Innovación/Agencia Estatal de Investigación MCIN/AEI/10.13039/501100011033.
\end{acknowledgements}

   \bibliographystyle{aa} 
  \bibliography{ref.bib} 

\begin{appendix}

\end{appendix}

\end{document}